\documentclass[fleqn,10pt]{wlscirep}
\usepackage[utf8]{inputenc}
\usepackage[T1]{fontenc}
\usepackage{longtable}
\usepackage{tabularx}
\usepackage{multirow}

\newcolumntype{A}{>{\centering\arraybackslash}p{3cm}}
\newcolumntype{B}{>{\centering\arraybackslash}p{2.5cm}}
\newcolumntype{C}{>{\centering\arraybackslash}p{1.5cm}}
\newcolumntype{D}{>{\centering\arraybackslash}p{2cm}}
\newcolumntype{E}{>{\centering\arraybackslash}p{2.5cm}}
\newcolumntype{F}{>{\centering\arraybackslash}p{3.5cm}}
\newcolumntype{G}{>{\centering\arraybackslash}p{3.5cm}}
\newcolumntype{H}{>{\centering\arraybackslash}p{2.8cm}}
\newcolumntype{I}{>{\centering\arraybackslash}p{2.75cm}}
\newcolumntype{J}{>{\centering\arraybackslash}p{1.25cm}}
\newcolumntype{K}{>{\centering\arraybackslash}p{1cm}}
\newcolumntype{L}{>{\centering\arraybackslash}p{1.75cm}}

\title{Quantifying gender imbalance in East Asian academia: Research career and citation practice}

\author[1, 2]{Kazuki Nakajima}
\author[2]{Ruodan Liu}
\author[3]{Kazuyuki Shudo}
\author[2, 4, 5, *]{Naoki Masuda}

\affil[1]{Department of  Mathematical  and Computing Science, Tokyo  Institute  of Technology,  Meguro-ku, Tokyo  152-8552, Japan.}
\affil[2]{Department  of  Mathematics,  State  University  of  New  York  at  Buffalo,  Buffalo  14260,  USA.}
\affil[3]{Academic Center for Computing and Media Studies, Kyoto University, Sakyo-ku, Kyoto 606-8501, Japan.}
\affil[4]{Computational  and Data-Enabled  Science  and  Engineering  Program,  State  University  of  New  York  at  Buffalo,  Buffalo  14260, USA.}
\affil[5]{Center for Computational Social Science, Kobe University, Kobe 657-8501, Japan.}
\affil[*]{Corresponding author: naokimas@buffalo.edu}

\begin{abstract}
Gender imbalance in academia has been confirmed in terms of a variety of indicators, and its magnitude often varies from country to country.
Europe and North America, which cover a large fraction of research workforce in the world, have been the main geographical regions for research on gender imbalance in academia. However, the academia in East Asia, which accounts for a substantial fraction of research, may be exposed to strong gender imbalance because
Asia has been facing persistent and stronger gender imbalance in society at large than Europe and North America.
Here we use publication data between 1950 and 2020 to analyze gender imbalance in academia in China, Japan, and South Korea in terms of the number of researchers, their career, and citation practice.
We found that, compared to the average of the other countries, gender imbalance is larger in these three East Asian countries in terms of the number of researchers and their citation practice and additionally in Japan in terms of research career. 
Moreover, we found that Japan has been exposed to the larger gender imbalance than China and South Korea in terms of research career and citation practice.
\end{abstract}

\keywords{science of science, research career, citation practice}

\begin{document}

\flushbottom
\maketitle
\thispagestyle{empty}

%\noindent Please note: Abbreviations should be introduced at the first mention in the main text \UTF{2013} no abbreviations lists. Suggested structure of main text (not enforced) is provided below.

\section{Introduction}

Gender imbalance persists in many aspects of society. It is present in academia in terms of research performance \cite{xie1998, besselaar2016}, academic awardees \cite{lincoln2012, lunnemann2019, meho2021, card2023}, faculty hiring \cite{steinpreis1999, mossracusin2012, sheltzer2014}, citations received by researchers \cite{caplar2017}, funding applications \cite{bornmann2007, witteman2019}, authorship positions \cite{lariviere2013, holman2018}, and many more \cite{llorens2021}.
Furthermore, it has been observed that female researchers are more disadvantaged than male researchers in hiring and promotion processes \cite{steinpreis1999, mossracusin2012, sheltzer2014, depaola2015} and that females tend to be less credited than males in research contributions \cite{ross2022}.
Mechanisms of these gender imbalances are intricately interrelated, complicating the formulation and evaluation of policies to address gender imbalance in academia \cite{llorens2021}.
Persistent gender imbalance may generate a naive interpretation that female and male researchers are not equivalent in their research capabilities, which may hamper women's progression in academia and undervalue women's academic contributions.
A growing body of evidence, however, suggests that the gender imbalance may arise from academic systems (e.g., research career \cite{holliday2014, evers2015, lerchenmueller2018, jadidi2018, huang2020, liang2020, bradshaw2021, morgan2021, kim2022, esther2023}) and practices of individual researchers (e.g., citation practice \cite{cameron2016, caplar2017, king2017, dion2018, nunkoo2019, dworkin2020, wang2021, teich2022, kong2022, tekles2022, kristina2022}) rather than from their research performance.

Gender imbalance in academia has been investigated in terms of various indicators. First, it has widely been quantified by the number of researchers.
The number of male researchers has been consistently larger than that of female researchers in various research disciplines \cite{holman2018, paswan2020, huang2020, chan2020}, across different career stages (e.g., assistant professors or lecturers \cite{auriol2020}, full professors \cite{barabino2020}), editorial boards \cite{hafeez2019, palser2022, liu2023}, and academic honorees and awardees \cite{lincoln2012, lunnemann2019, meho2021, card2023}.
Second, gender imbalance in academia has been investigated in terms of individual researchers' careers \cite{holliday2014, evers2015, lerchenmueller2018, jadidi2018, huang2020, liang2020, bradshaw2021, morgan2021, kim2022, esther2023}.
For example, gender imbalance in the likelihood of becoming a principal investigator is largely explained by that in the number of publications in the early career in life sciences \cite{lerchenmueller2018}. Another study showed that gender imbalance in the number of publications in the entire career is largely explained by that in the publishing career length \cite{huang2020}.
Third, citation practices are subject to gender imbalance~\cite{cameron2016, caplar2017, king2017, dion2018, nunkoo2019, dworkin2020, wang2021, teich2022, kong2022, tekles2022, kristina2022}.
For example, male researchers tend to cite their own papers more often than female researchers do \cite{cameron2016, king2017}. As another example, male researchers tend to cite more papers written by male than female researchers with the difference in the number of papers written by male and female authors being taken into account \cite{dworkin2020, wang2021, teich2022}.
Because the number of citations received by a researcher is a basic indicator of their research performance, gender imbalance in terms of the number of citations received may be associated with other facets of gender imbalance such as research career.
These and other data and analysis results are expected to promote the formulation of policies to improve the gender imbalance in academia \cite{llorens2021, casad2021}. In this study, we focus on the aforementioned three topics, i.e., the number of researchers, research careers, and citation practices, among many topics.

In general, gender imbalance in society differs from country to country. Evidence suggests that Asia has been exposed to larger gender imbalance than Europe and North America.
The World Economic Forum (WEF) in 2022 reported that Europe and North America each has ranked higher than Central Asia, ``East Asia and the Pacific'', and South Asia between 2006 and 2022 in terms of the gender gap index \cite{wrf_report_2022}.
Therefore, gender imbalance in academia in Asia may be larger than that in Europe and North America. 
However, many previous studies (e.g., Refs.~\cite{abramo2013, morgan2021, kwiek2021, abramo2021, portugal_gender_report_2021, kwiek2021_2}) have focused on researchers in Europe and North America for two main reasons.
First, researchers in Europe and North America have been mainly leading advanced research and cover a large fraction of workforce.
For example, the publications from Europe, North America, and Asia received 35.3\%, 42.3\%, and 17.7\%, respectively, of the approximately 62 million citations made by over eight million publications published between 2003 and 2010 \cite{pan2012}.
In addition, 267 Nobel laureates in Physiology or Medicine, Physics, or Chemistry until 2017 are from Europe, 310 from North America, and only 26 from the Asia-Pacific \cite{heinze2019}.
Second, in general, it is difficult to infer the gender from Asian names \cite{santamaria2018}.
For this reason, a previous study on gender imbalance made a conservative decision to exclude researchers working in Asian countries (i.e., China, Japan, Malaysia, North Korea, Singapore, and South Korea) from the analysis \cite{huang2020}.

We focus on gender imbalance in academia in China, Japan, and South Korea because they are globally research-active East Asian countries.
Recent research outputs from these three countries are comparable to those from Europe and North America.
For example, the rapid growth of China's share of research outputs in the world is remarkable \cite{zhou2006, gomez2022}.
Japanese research highly ranked in the world until about 2000, whereas its research capability has been declining in recent years \cite{nistep_2022}. 
South Korea has ranked high in the world in terms of research and development expenditure as a percentage of the gross domestic product \cite{r_and_d_2022} and ranks the sixth of 132 countries in the global innovation index in 2022 \cite{gii_2022}.
Additionally, Times Higher Education reported that the three countries hold 19 of the top 200 universities listed in the world university ranking 2023 (11 universities in China, two in Japan, and six in South Korea) \cite{world_univ_rank_2023}. 
These numbers have largely been growing over recent years, while European and North American universities are yet dominant in this ranking (e.g., 58 universities in the US, 28 in the UK, and 22 in Germany) and other rankings.
While Singapore also has a high research capability (e.g., it holds two universities that ranked 19th and 36th in the same ranking), we exclude Singapore because its small number of researchers due to the country size would make our data analysis difficult.

In the present study, we hypothesize that China, Japan, and South Korea are exposed to larger gender imbalance than other research-active countries.
These countries in East Asia have long faced gender imbalance across a multitude of societal aspects \cite{molony2016}.
In fact, China, Japan, and South Korea ranked 102nd, 116th, and 99th, respectively, out of 146 countries in the global gender gap in 2022 \cite{wrf_report_2022}.
There are a number of previous studies on historical factors associated with gender imbalance in East Asian societies \cite{raymo2015}.
Furthermore, recent studies provided evidence of gender imbalance in academia in these three East Asian countries, while it was only in terms of the number of researchers \cite{lariviere2013, chan2020, meho2021}.
For example, all the three countries are included in the bottom 10 among 43 countries in terms of the proportion of female researchers among the most highly cited researchers \cite{chan2020}.
These observations motivate us to further compare gender imbalance among the three East Asian countries and the other countries.

We also hypothesize that Japanese academia is exposed to larger gender imbalance than Chinese and South Korean academia.
Japan has been exposed to large gender imbalance across areas of society including industries \cite{wright1995}, education \cite{mcdaniel2010}, and politics \cite{pharr1981}.
In fact, Japan ranks the lowest in the global gender gap in 2022 among the G7 countries as well as in the region of ``East Asia and the Pacific'' \cite{wrf_report_2022}.
Moreover, a previous study reported that women accounted for 31 percent of doctoral graduates in Japan in 2014, where the percentage is the lowest among the 14 countries investigated including China and South Korea \cite{shin2018}.
Such a scarcity of female students is a facet of gender imbalance in Japanese academia.
In addition, previous studies suggested that Japan has a larger gender imbalance in terms of the number of researchers than other countries including China and South Korea \cite{lariviere2013, holman2018}.
However, gender imbalance among the three East Asian countries is not still well understood beyond in terms of the number of researchers.
A technical barrier here is to collect sufficient samples of gender-assigned researchers in the three East Asian countries based on their names \cite{bendels2018, gender_report_2020, huang2020}. 

To test these hypotheses, we use publication data between 1950 and 2020 provided by the OpenAlex project \cite{priem2022} and quantify gender imbalance in the number of researchers, research career, and citation practice of researchers in different countries including the three East Asian countries.
To this end, we use samples of 18,002,917 gender-assigned researchers, with 387,925 researchers in China, 1,267,205 researchers in Japan, 342,452 researchers in South Korea, and 16,005,335 researchers in the other countries.
We discuss limitations of the samples used in our analyses in Section \ref{section:4}.

\section{Methods}

\subsection{Data}

We use publication data from a snapshot of OpenAlex released on July 9th, 2022.
OpenAlex is an open database of research publication records, which replaces the discontinued Microsoft Academic Graph (MAG) \cite{priem2022}.
See Section \ref{section:4} for potential problems with this data set.
We used the 104,169,109 papers that were published between 1950 and 2020 because the number of academic authors (i.e., those who write academic papers) working in South Korea before 1950 is small. For each paper $z$, the publication date, ``concepts'', which are research fields of the paper, publication type (e.g., ``Journal Article'' and ``Proceedings Article''), authors' IDs, papers cited by $z$, and author position (i.e., first, middle, or last) are available.
We assigned one of the 19 research disciplines to each paper $z$ based on the concepts associated with $z$ (see Supplementary Section S1 for the assignment method).

\subsection{Data preprocessing}

We will compare gender imbalance in the research career and citation practice among authors working in China, Japan, and South Korea.
To this end, we infer the gender and nationality of authors from the publication data in various countries. 

\subsubsection{Identifying unique authors}

Author-name disambiguation is a challenging problem in bibliographic data analysis \cite{ferreira2012}.
Like the MAG, the OpenAlex uses a proprietary algorithm to identify and assign a unique ID to each author using their name, publication record, citation patterns, and if available, Open Researcher and Contributor ID (ORCID) \cite{priem2022}.
The OpenAlex's author-name disambiguation algorithm identified 40,134,800 authors.
We assigned one of the 19 research disciplines to each of the 37,921,048 authors (94.5\%) based on their papers; we assigned no discipline to the other authors (see Supplementary Section S1 for the assignment method).
We also assigned a country to each author based on the country code (i.e., the ISO two-letter country code) associated with their institutions (see Supplementary Section S2 for the assignment method). 
As a result, 7,684,496 authors were affiliated with an institution in China, 2,103,087 authors in Japan, 784,120 authors in South Korea, and 29,563,097 authors in other countries. It should be noted that this country assignment is distinct from the author's country of origin, their ethnicity, or the language that their name is from.

\subsubsection{Separating first and last names}

We use the first name to infer an author's gender. 
However, the author's name is not explicitly separated into the first and last names in the OpenAlex data.
Therefore, we enumerated candidates of their first and last names as follows. 
First, we excluded the 112,763 authors (0.3\%) with a one-word name, which is presumably due to error in the original data.
Then, because different words in the author's name are separated by space (e.g., John Smith),
we regarded that the last name is the last space-separated word.
For example, the last name of ``John Smith'' is Smith according to our procedure.

Next, we acquired the candidates of the first name for the authors in China, Japan, or South Korea (i.e., those for whom we estimated the country to be China, Japan, or South Korea from their affiliations) as follows.
Chinese or Japanese names written in English typically consist of two words separated by space (e.g., Meiling Jiang or Hitomi Yamada).
In South Korean names, the first two or more space-separated words may be their first name. 
For example, ``Gil Dong'' may be the first name of the name ``Gil Dong Hong''. 
Therefore, if the author is in China, Japan, or South Korea and their name consists of $k$ space-separated words, denoted by $w_1, w_2, \ldots, w_k$, then we use $w_1$, $w_1 w_2$, $\ldots$, and $w_1 \cdots w_{k-1}$ as $k-1$ candidates of their first name. 
For example, the unique candidate of the first name for the name ``Gildong Hong'' is ``Gildong''. In contrast, the candidates of the first name for the name ``Gil Dong Hong'' are ``Gil'' and ``Gil Dong''.

For the rest of the authors, i.e., those who reside in countries other than China, Japan, and South Korea, we assume that the first space-separated word of their name is their first name.

\subsubsection{Gender of authors}

We infer an author's gender using an application programming interface (API) named the Gender API\footnote{\url{https://gender-api.com/en/} (Accessed December 2022)}.
The database constituting Gender API contains over six million unique names across 191 countries.
Gender API has been deployed in previous studies that investigated gender imbalance in academia~\cite{caplar2017, dworkin2020, squazzoni2021, teich2022}. 
Furthermore, its accuracy for Asian names is superior to competing services on a benchmark data set \cite{santamaria2018}.

We constructed the gender assignment methods that yield at least 90\% classification accuracy on the small set of samples with the ground truth, which we call the test set, as follows.
First, when one inputs a first-name candidate to Gender API, it returns either ``female'', ``male'', or ``unknown'', its `accuracy' (which is a terminology of Gender API and different from the classification accuracy on the test set discussed below; therefore we put the quotation marks here and in the following text), and the number of samples of the provided first name in the database.
Second, for the authors in either China, Japan, or South Korea, we decided to feed their country, which is an optional input to the API, in addition to each candidate of their first name. 
In contrast, for the authors in the other countries, we fed each candidate of their first name to the API without specifying their country. 
We used the country in some cases and not others because the classification accuracy on the test set was higher in this way.
Note that we assessed the classification accuracy of any gender assignment method by manually checking the correctness of the assigned gender on the test set (see Supplementary Section S3 for the details).
Third, we discarded the API's output when the `accuracy' that the API returned was smaller than $\theta$ or the number of samples was smaller than $n_s$.
We then identified the largest `accuracy' value among all the input first-name candidates that returned female as output.
Similarly, we identified the largest `accuracy' value among all the input first-name candidates that returned male as output.
If the former was larger than the latter, we assigned female to the author.
If the former was smaller than the latter, we assigned male to the author.
If the former was equal to the latter, then we did not assign female or male to the author. 

We set the $\theta$ and $n_s$ values depending on a given author's country and first publication year. 
Specifically, for the authors in China, we set $\theta =$ 90\% and $n_s=$ 10 if their first publication year, denoted by $y$, is 1990 or before; $\theta =$ 99\% and $n_s=$ 10 if $1991 \le y \le 2000$; $\theta =$ 99\% and $n_s=$ 50 if $2001 \le y \le 2010$; $\theta=$ 95\% and $n_s=$ 10 if $2011 \le y \le 2020$. For the authors in Japan, we set $\theta =$ 99\% and $n_s=$ 10 if $y\le 1990$, and $\theta =$ 90\% and $n_s=$ 10 if $1991 \le y \le 2020$. 
For the authors in the other countries including South Korea, we set $\theta =$ 90\% and $n_s=$ 10 for any $y$. With this choice, the classification accuracy on the test set is at least 90\%
for each pair of the country (i.e., China, Japan, South Korea, and the other countries) and
the year group (i.e., (i) $y \le 1990$, (ii) $1991 \le y \le 2000$, (iii) $2001 \le y \le 2010$, and (iv) $2011 \le y \le 2020$). 

In this manner, we assigned a binary gender to 18,002,917 (with 6,485,368 females) out of the 40,134,800 authors, with 387,925 authors (with 89,838 females) in China, 1,267,205 authors (with 216,579 females) in Japan, 342,452 authors (with 44,257 females) in South Korea, and 16,005,335 authors (with 6,134,694 females) in the other countries. 
The fractions of gender-assigned authors are 5.0\% ($\approx$ 387,925/7,684,496), 60.2\% ($\approx$ 1,267,205/2,103,087), 43.7\% ($\approx$ 342,452/784,120), and 54.1\% ($\approx$ 16,005,335/29,563,097) in China, Japan, South Korea, and the other countries, respectively.
The fraction of gender-assigned authors is low for our Chinese samples. Therefore, our Chinese samples may lack representativeness compared to the samples from Japan, Korea, and the other countries. We discuss this limitation in Section \ref{section:4}.
The authors in the top 31 countries in terms of the number of gender-assigned authors combined account for 90.0\% of those in the other countries. In descending order in terms of the number of gender-assigned authors, these countries are US, UK, Germany, Brazil, France, India, Canada, Spain, Italy, Australia, Islamic Republic of Iran, Netherlands, Switzerland, Mexico, Russian Federation, Poland, Turkey, Sweden, Indonesia, Belgium, Egypt, Colombia, Israel, Denmark, Austria, Pakistan, Czechia, Greece, Argentina, Finland, and Malaysia.

\subsubsection{Nationality of authors}

A Japanese author and a foreign author in Japan may be subject to different degrees of gender imbalance. 
To examine this possibility, we used an API named Nationalize.io\footnote{\url{https://nationalize.io/} (Accessed December 2022)} to infer the nationality of the authors in China, Japan, or South Korea.
For a given first or last name, Nationalize.io returns the candidates of the nationality of the person and their probabilities.
The database constituting Nationalize.io contains approximately 114 million names across many countries.

For each author $u$, we ran Nationalize.io separately with $u$'s last name as input and with each candidate of $u$'s first name as input. 
For each input, Nationalize.io returned up to five nationality candidates with the largest probabilities.
Denote by $c$ the country in which $u$ resides (i.e., China, Japan, or South Korea). If Nationalize.io outputs $c$ as a candidate of $u$'s nationality and its probability is at least 0.9 for at least one candidate of $u$'s first name or $u$'s last name, we set $c$ as $u$'s nationality and therefore regard that $u$ is a native author (i.e., those who work in the country where they have nationality) in country $c$.
In contrast, we regard that $u$ is a non-native author if the probability of country $c$ is not greater than 0.1 for all candidates of $u$'s first name and $u$'s last name.
Specifically, we do so if at least one of the following three criteria is met for each of the author's name words as input: (i) Nationalize.io outputs $c$ as a candidate of the nationality and its probability is 0.1 or smaller; (ii) Nationalize.io does not output $c$ as a candidate of the nationality, and the smallest probability of the country in the output of Nationalize.io is at most 0.1, or (iii) Nationalize.io does not output $c$ as a candidate, and the sum of the probabilities associated with all the output countries is at least 0.9.
Note that, if criterion (ii) or (iii) is met, then the probability of nationality $c$ is at most 0.1 because the API returns up to five nationality candidates with the largest probabilities.
Otherwise, we regarded the author to be neither native nor non-native to the country.

For example, if the author's name is ``Hitomi Yamada'', we separately input ``Hitomi'' and ``Yamada'' to Nationalize.io. Suppose that Nationalize.io returns the following nationalities and their probabilities: $\{$(Japan, 0.80), (China, 0.05)$\}$ for ``Hitomi'' and $\{$(Japan, 0.91), (Thailand, 0.02), (Vietnam, 0.01)$\}$ for ``Yamada''. 
If the author is in Japan, we regard that the author is a Japanese native because the probability of Japan given ``Yamada'' is larger than $0.9$.
In contrast, if Hitomi Yamada is in China, we regard that the author is non-native because the output for the input name ``Hitomi'' satisfies criterion (i) and that for the input name ``Yamada'' satisfies criteria (ii) and (iii).

Among the gender-assigned authors, we found 131,372 native and 13,700 non-native authors in China, 1,179,036 native and 17,030 non-native authors in Japan, and 200,325 native and 6,338 non-native authors in South Korea. 
Note that many authors are not assigned to be native or non-native.

\subsection{Characterizing gender imbalance in research career}

We characterize the career of author $u$ using the five indicators proposed in Ref.~\cite{huang2020}: (i) total productivity defined as the number of papers authored by $u$; (ii) total impact defined as the sum of the citation impact across papers authored by $u$ (see Supplementary Section S4 for the definition of citation impact of a paper); (iii) career length defined as the difference between the publication dates of the first and last papers authored by $u$, which we divide by 365; (iv) annual productivity defined as the ratio of $u$'s total productivity to $u$'s career length; (v) number of coauthors defined by the number of authors that coauthored at least one paper with $u$.

For the combination of an indicator and a set of authors, we define the gender gap by $(\mu_{\text{F}} - \mu_{\text{M}})/\mu_{\text{M}}$, where $\mu_{\text{F}}$ and $\mu_{\text{M}}$ denote the mean value of the indicator for the female and male authors in the set, respectively.
A positive gender gap value implies that females have stronger careers than males in terms of the indicator used, and vice versa.

\subsection{Characterizing gender imbalance in citation practice}

We quantify gender imbalance in citation practice of authors in China, Japan, South Korea, and the other countries using the methods developed in Ref.~\cite{dworkin2020}.
We focus on citations between pairs of papers among the 27,616,941 papers for each of which both the first and last authors have the same country, including the case of a country other than China, Japan, and South Korea.
We denote by $S$ the set of all the papers.
We identify four categories, i.e., MM, MW, WM, and WW, based on the gender of the first and last authors, where the first letter, M or W, indicates that the first author is male (man) or female (woman), respectively, and the second letter indicates the gender of the last author. We classify single-author papers to the MM or WW category.

Consider two subsets of papers denoted by $S_{\text{from}} \subseteq S$ and $S_{\text{to}} \subseteq S$. We measure the extent to which the papers in $S_{\text{from}}$ over- or under-cite the papers in $S_{\text{to}}$ and in a given gender category (i.e., MM, MW, WM, or WW).
To this end, for each paper $z \in S_{\text{from}}$, we focused on citations made by $z$ to each paper $z' \in S_{\text{to}}$ such that (i) the publication date of $z'$ is at most ten years older than that of $z$ and (ii) both first and last authors of $z'$ are neither the first nor last author of $z$. 
The original method used all the citations made by $z$ to any papers published before $z$ to investigate gender imbalance in citation practice \cite{dworkin2020}. 
In contrast, a previous study adopted criterion (i) to measure gender imbalance in the research career \cite{huang2020}. We adopt criterion (i) because in this manner we can simultaneously examine gender imbalance in research career and citation practice with the same data. We impose criterion (ii) to avoid self-citations. 
We denote by $\overline{S}_{\text{to}}(z)$ the set of the papers that meet criteria (i) and (ii) for a given paper $z$.

Given a gender category $g \in \{$MM, MW, WM, WW$\}$, we first count the citations made by each paper $z \in S_\text{from}$ to the papers in $\overline{S}_{\text{to}}(z)$ and in gender category $g$, which we denote by $n_{z\to g, \text{obs}}$. 
Then, we compare $\sum_{z \in S_\text{from}} n_{z\to g, \text{obs}}$ with the expectation under the so-called random-draws model \cite{dworkin2020}.
The random-draws model assumes that $c_z$ papers cited by paper $z \in S_{\text{from}}$ are drawn from $\overline{S}_{\text{to}}(z)$ uniformly at random, where $c_z$ is the number of citations made by $z$ to the papers in $\overline{S}_{\text{to}}(z)$. 
Then, the expected number of citations that the papers in $\overline{S}_{\text{to}}(z)$ and in gender category $g$ receive from $z$ under the random-draws model, denoted by $n_{z \to g, \text{rand}}$, is equal to $c_z p_{z,g}$, where $p_{z,g}$ is the fraction of the papers that are in both $\overline{S}_{\text{to}}(z)$ and $g$.
We sum $n_{z \to g, \text{rand}}$ over $z$, which yields the expected number of citations received by the papers in both $S_{\text{to}}$ and $g$.
Then, we calculate the over/under-citation of the papers that belong to both $S_{\text{to}}$ and $g$ as $(\sum_{z \in S_\text{from}} n_{z\to g, \text{obs}} - \sum_{z \in S_\text{from}} n_{z\to g, \text{rand}})/\sum_{z \in S_\text{from}} n_{z\to g, \text{rand}}$.

Previous studies additionally deployed another reference model, called the relevant characteristics model, for calculating the expected number of citations \cite{dworkin2020}. 
We do not use this method because the Newton's method to determine the smoothing parameters does not sufficiently converge due to a huge number of papers.
Therefore, the present study reports the over/under-citation obtained using the random-draws model.

\section{Results}

\subsection{Gender imbalance in research career}

\begin{figure}
  \begin{center}
	\includegraphics[scale=0.158]{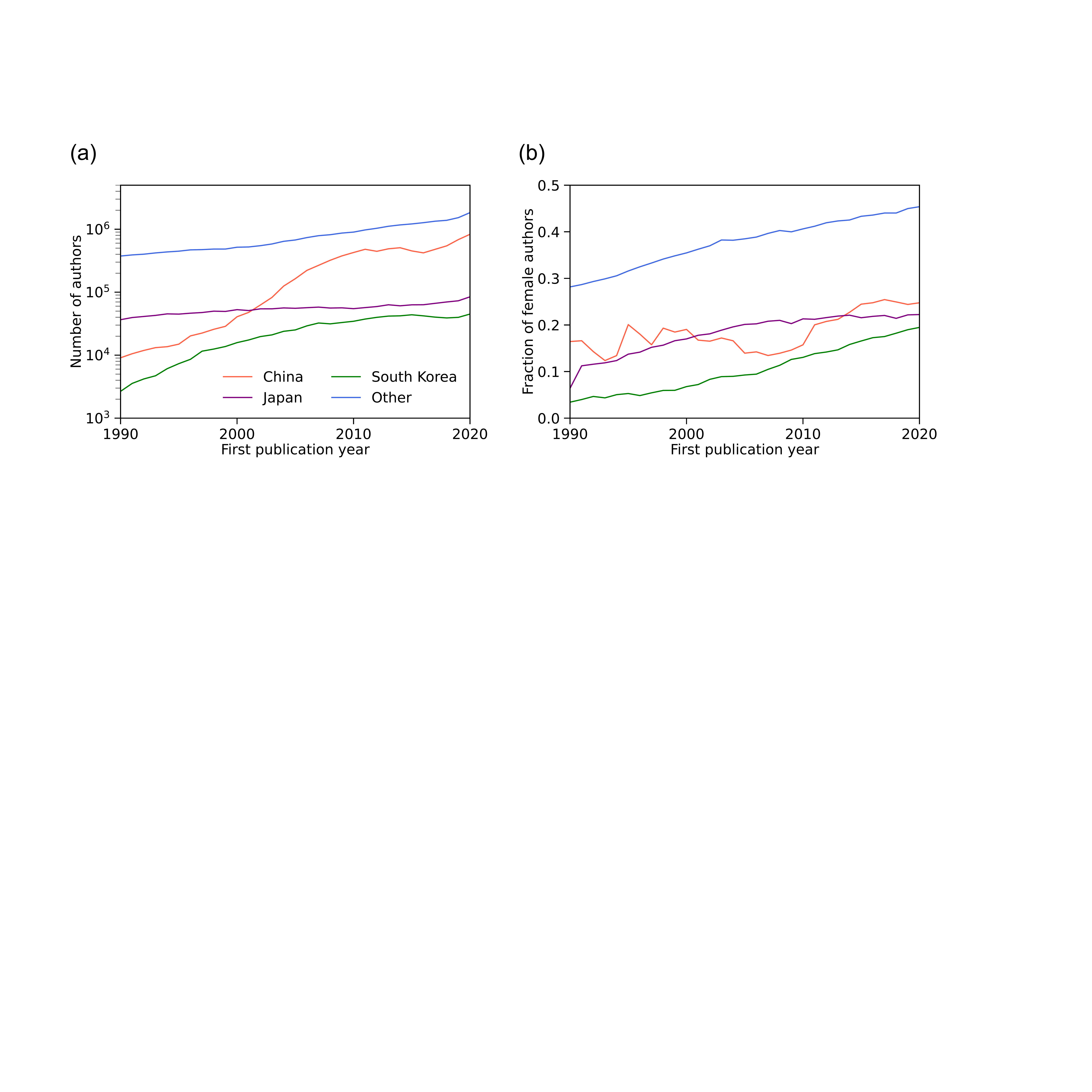}
  \end{center}
  \caption{Counts of authors in China, Japan, South Korea, and the other countries. (a) Number of authors between 1990 and 2020. (b) Fraction of female authors between 1990 and 2020.
}
  \label{fig:1}
\end{figure}

We first examine gender imbalance in the set of authors in China (i.e., their country of affiliation was estimated to be China), Japan, South Korea, and the other countries.
We identified 7,684,496 authors in China, 2,103,087 authors in Japan, 784,120 authors in South Korea, and 29,563,097 authors in the other countries.
In each country group, the number of authors has increased from 1990 to 2020, in particular in China (see Fig.~\ref{fig:1}(a)); by definition, we counted the authors who had published their first paper in each given year. 
Among these authors, we were able to assign a binary gender to 387,925 authors in China, 1,267,205 authors in Japan, 342,452 authors in South Korea, and 16,005,335 authors in the other countries.
The fractions of female authors are 23.2\%, 17.1\%, and 12.9\% in China, Japan, and South Korea, respectively. The ranking among the three East Asian countries in our data set is consistent with previous results \cite{chan2020, meho2022}.
These fractions are smaller than that for the other countries in our data set (i.e., 38.3\%). As a reference, the fraction for 83 countries excluding China, Japan, and South Korea was 28.5\% in a previous study \cite{huang2020}.
The fraction of female authors is the highest in psychology and the lowest in engineering among the 19 disciplines in the three East Asian countries, whereas it is the highest in sociology and the lowest in physics in the other countries (see Supplementary Section S5 for details).
The fraction of female authors has increased over the three decades in each country group (see Fig.~\ref{fig:1}(b)). 
Specifically, while female authors represented 16.4\%, 6.4\%, 3.4\%, and 28.2\% in China, Japan, South Korea, and the other countries, respectively, in 1990, the fraction increased to 24.8\%, 22.2\%, 19.5\%, and 45.4\% in 2020. 
The fractions for China, Japan, and South Korea have been smaller than that for the other countries over the three decades (see Fig.~\ref{fig:1}(b)).

\begin{figure}
  \begin{center}
	\includegraphics[scale=0.21]{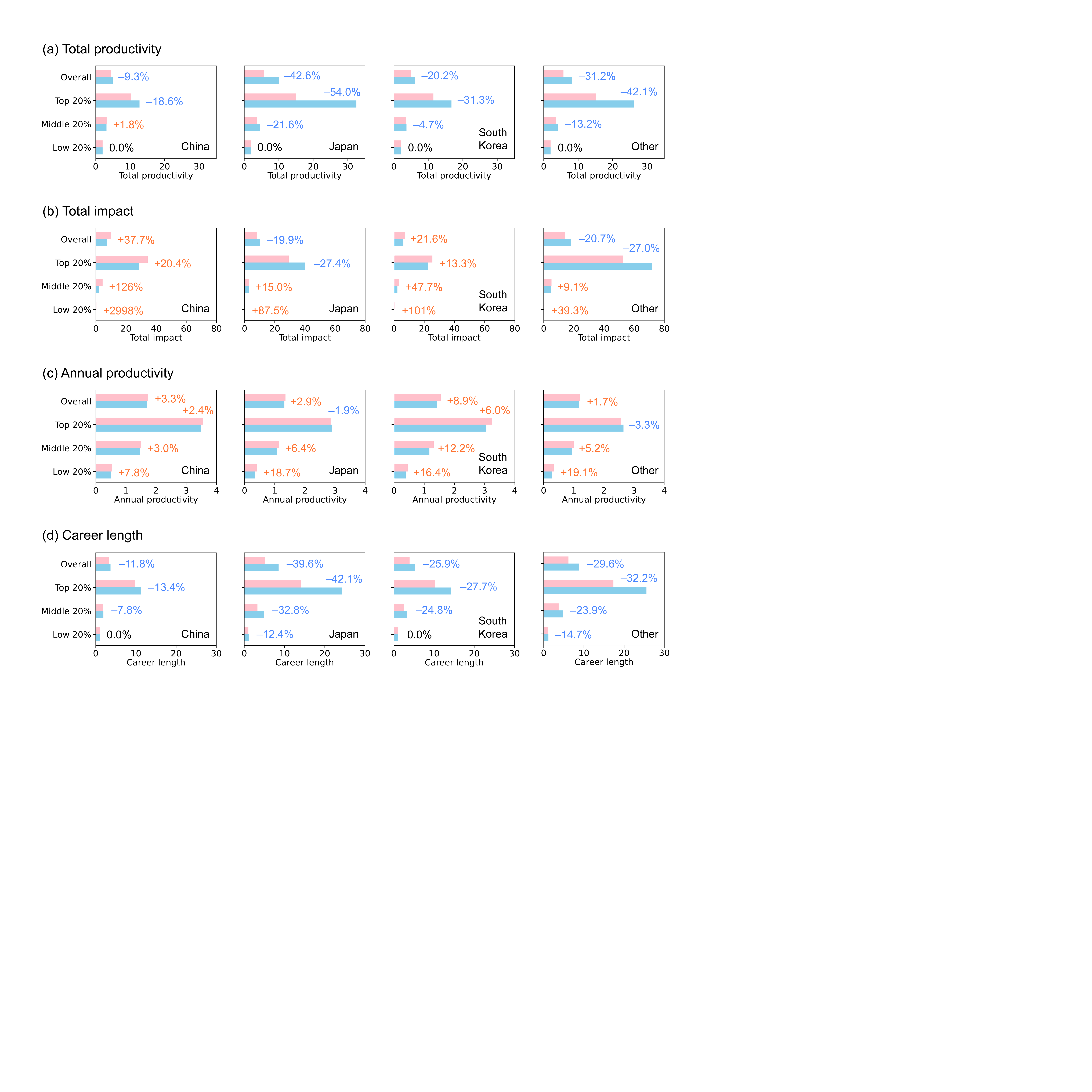}
  \end{center}
  \caption{Gender imbalance in the research career for the authors in China, Japan, South Korea, and the other countries. (a) Total productivity. (b) Total impact. (c) Annual productivity. (d) Career length. In each panel, the gender gap for all authors in the country group, top 20\%, middle 20\%, and bottom 20\% authors are shown. For example, the top 20\% refers to the gender gap between the top 20\% females and the top 20\% males in the country group.}
  \label{fig:2}
\end{figure}

We now compare gender imbalance in the research career among different countries.
In the remainder of this subsection, we focus on the gender-assigned authors in each country group (i.e., China, Japan, South Korea, or the other countries) who meet the following criteria: (i) they publish at least two papers, (ii) the difference between the publication dates of their first and last papers is more than 365 days, (iii) they publish fewer than 20 papers per year on average, and (iv) their last publication date is on or before December 31st, 2015. 
Criteria (i)--(iii) are the same as those employed in Ref.~\cite{huang2020}. 
We extended the year of the last publication in criterion (iv) from 2010, which was used in Ref.~\cite{huang2020} given that the last year that their data set covered was 2016, to 2015 given that the last year that our data set covers is 2020.
There are 17,813 such authors (with 20.3\% being females) in China, 327,946 (14.0\% females) in Japan, 57,466 (9.8\% females) in South Korea, and 2,544,059 (33.4\% females) in the other countries.

We have found that the authors in Japan are exposed to a notably larger gender gap in terms of the total productivity than those in ``the other countries'' group, whose gender gap is larger than those in China and South Korea (see the bars labeled ``Overall'' in Fig.~\ref{fig:2}(a)).
These results persist when we compare the top $20\%$ female authors and the top $20\%$ male authors in each country group (see the bars labeled ``Top 20\%'' in Fig.~\ref{fig:2}(a)).
When we compare the middle 20\% females and the middle 20\% males in each country group, the gender gap is almost eliminated for China and South Korea but remained for Japan and the other countries (see the bars labeled ``Middle 20\%'' in Fig.~\ref{fig:2}(a)).
Last, there is no gender gap between the bottom 20\% females and the bottom 20\% males in each country group (see the bars labeled ``Low 20\%'' in Fig.~\ref{fig:2}(a)).
Our results for China and South Korea including the magnification of the gender gap in the ``Top 20\%'' authors and its suppression in the middle and bottom groups of authors are similar to those in a recent study analyzing 83 countries although that study excluded China, Japan, and South Korea \cite{huang2020}. 
All the non-zero differences between the two genders shown in Fig.~\ref{fig:2}(a) are significant (see Supplementary Section S6 for statistical results).

In terms of the total impact of an author, we have observed qualitatively different results (see Fig.~\ref{fig:2}(b)).
Female authors receive more citations than male authors on average in China and South Korea and vice versa in Japan and the other countries.
These results remain qualitatively the same when we compare the top 20\% females and the top 20\% males in each country group.
In contrast, in the comparison between the middle 20\% females and the middle 20\% males, and between the bottom 20\% females and the bottom 20\% males, the females receive more citations than the males in each country group. All these gender gaps are statistically significant (see Supplementary Section S6).
Our results for China and South Korea are different from the previous results \cite{huang2020} in that the overall and top 20\% females receive more citations than the males in China and South Korea in ours. Our results for Japan and the other countries are consistent with the previous results \cite{huang2020}.

For further investigation of the gender gap in total productivity, we decompose the total productivity of each author into the product of their annual productivity and career length \cite{huang2020}; the career length is defined as the difference between the first and last publication dates.
We have found a small gender gap in the annual productivity of the authors in each country group (see Fig.~\ref{fig:2}(c)).
The gender gap remains small in the comparison between the top 20\% females and the top 20\% males in each country group as well.
Consistent with these results, many of the gender gaps shown in Fig.~\ref{fig:2}(c) are statistically insignificant (see Supplementary Section S6).
These results are consistent with the previous results~\cite{huang2020}.

The lack of gender gap in annual productivity in each country group suggests that the country-to-country difference in the gender gap in total productivity may be ascribed to that in career length.
As expected, the gender gap in the career length is present in each country group both when we compare all females and males and when we compare the top 20\% authors of each gender (see Fig.~\ref{fig:2}(d)). 
All the non-zero gender gaps shown in Fig.~\ref{fig:2}(d) are significant (see Supplementary Section S6).
The gender gap in the career length is the strongest in Japan, then ``the other countries'' group, South Korea, and the weakest in China. 
This result is also expected because the annual productivity is roughly free of the gender gap and the gender gap in the total productivity is also the strongest in Japan, then ``the other countries'' group, South Korea, and the weakest in China. 
The presence of a gender gap in career length for the overall and top 20\% authors by gender in each country group is consistent with the previous results \cite{huang2020}.

\begin{figure}
  \begin{center}
	\includegraphics[scale=0.21]{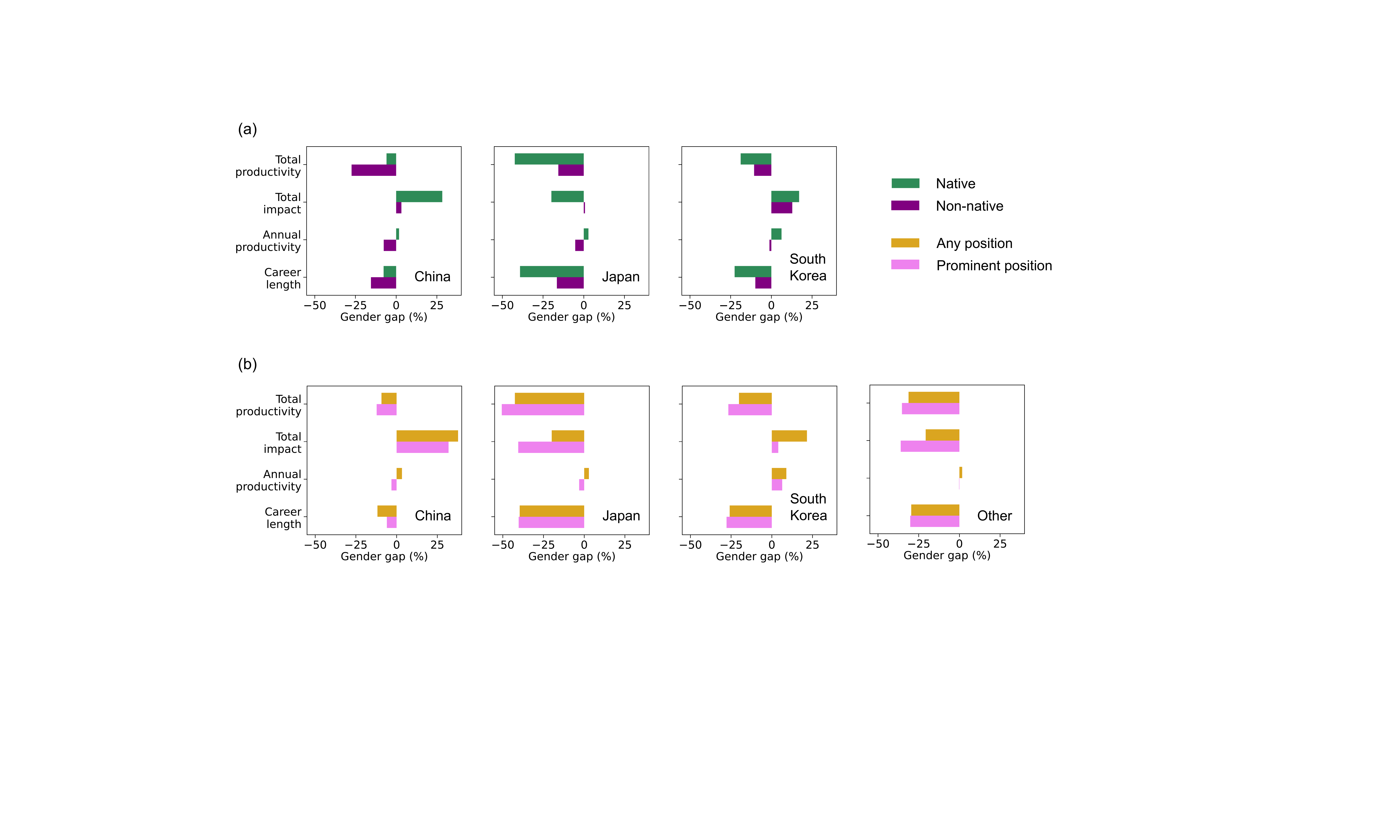}
  \end{center}
  \caption{Gender imbalance attributed to the author's nationality or author's position in the author list. (a) Native versus non-native authors. (b) Any eligible papers versus those in which the author occupies a prominent position.}
  \label{fig:3}
\end{figure}

To sum up, we found that the gender gaps in terms of the four indicators in each country group are overall similar to the previous results \cite{huang2020} and that these gender gaps are the largest in Japan, then ``the other countries'' group, South Korea, and weakest in China.

Gender imbalance may be different between native and non-native authors in a country. 
For example, Chinese authors working in China and foreign authors working in China may be exposed to different levels of gender imbalance.
Therefore, we compare gender imbalance between ``native'' authors and ``non-native'' authors in China, Japan, and South Korea.
There are 4,915 natives (with 20.7\% females) and 974 non-natives (with 25.6\% females) in China; 311,550 natives (with 13.2\% females) and 2,858 non-natives (with 19.0\% females) in Japan; and 32,144 natives (with 13.0\% females) and 666 non-natives (with 17.7\% females) in South Korea.
We compare the four indices of gender gap between native and non-native authors in China, Japan, and South Korea in Fig.~\ref{fig:3}(a).
The figure suggests that non-native females tend to be at more disadvantage than native females in China.
In contrast, in Japan, native females tend to be at more disadvantage in terms of research career than non-native females except in terms of annual productivity.
The results for South Korea are, roughly speaking, intermediate between those for China and Japan.
These results suggest that natives and non-natives perceive different gender imbalance and how they do so depends on countries even within East Asia.

While we have been ignorant of the author's position, a previous study suggested that female authors are less likely to appear in the prominent author position (i.e., the first- or last-author position or sole author) \cite{lariviere2013, west2013}; authors in these positions generally play main roles in the research and writing of the paper in many research disciplines.
To investigate whether gender imbalance depends on the author position, we now restrict the set of papers for each author $u$ to those in which $u$ is in the prominent position. Then, using the restricted set of papers, we examine whether or not $u$ meets the same four criteria for filtering the authors. Note that $u$ may not satisfy the criteria after we restrict the papers in this manner even if $u$ satisfies the same criteria based on the entire set of papers
(e.g., $u$ has written papers but never as prominent author).
After removal of the authors that do not satisfy the four criteria as prominent author,
there are 10,180 authors (with 15.4\% females) for China, 185,874 authors (with 10.2\% females) for Japan, 31,936 authors (with 8.3\% females) for South Korea, and 1,825,707 authors (with 29.4\% females) for the other countries.
The gender gap qualitatively remains the same in each country group when one calculates the four indicators of research career only using the papers in which $u$ is in the prominent position (see Fig.~\ref{fig:3}(b)). In other words, females publish fewer papers, receive more citations in China and South Korea but fewer citations in Japan and the other countries, and have shorter career lengths than males, and the annual productivity is similar between the females and males.
Furthermore, female authors tend to be at more disadvantage in the prominent author position than in any author position in all the country groups, except in terms of the career length for China (see Fig.~\ref{fig:3}(b)).

\begin{figure}[t]
  \begin{center}
	\includegraphics[scale=0.19]{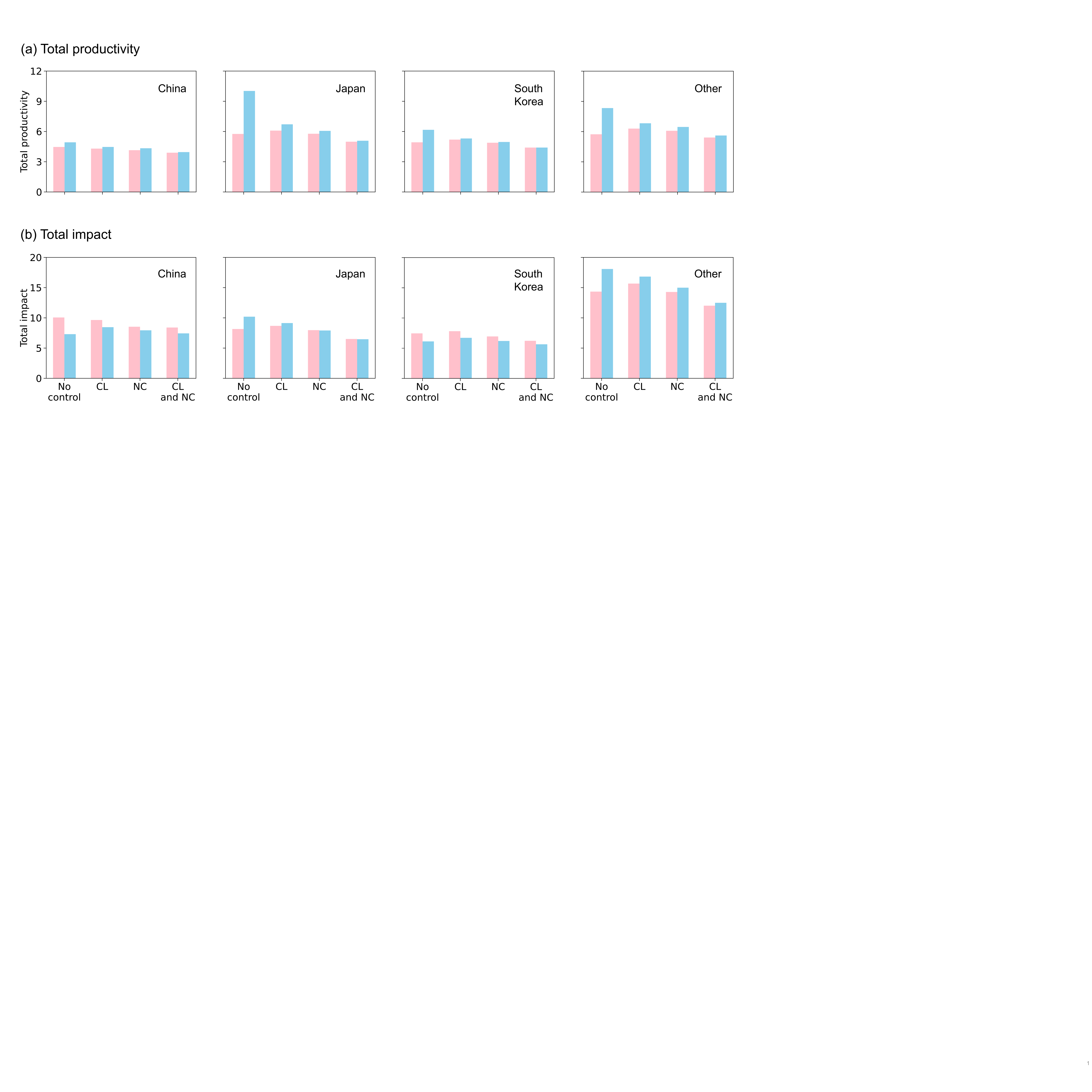}
  \end{center}
  \caption{Gender imbalance in research performance after controlling for the career length or the number of coauthors. (a) Total productivity. (b) Total impact. Each panel shows the results of the matching experiment in which the author's career length (CL) and/or their number of coauthors (NC) are controlled in addition to their country group, year of the first publication, and research discipline. The bar labeled ``No control'' in each panel is identical to that labeled ``Overall'' for the corresponding indicator and country group in Fig.~\ref{fig:2}.}
  \label{fig:4}
\end{figure}

Among our research performance indicators (i.e., total productivity, total impact, and annual productivity), we have observed gender imbalance in the total productivity and total impact.
Prior research showed that gender imbalance in the career length \cite{huang2020} and that in the number of coauthors \cite{li2022} explain most of the variance in the research performance.
To test the existence of similar correlates in each country group, we carry out matching experiments~\cite{huang2020, li2022} in which the gender gap in terms of the career length or number of coauthors is eliminated. We generate the three matched sets of authors as follows. 
For each female author, we select without replacement a male that has the same country group, the same year of the first publication, the same discipline, and the same career length in the case of the first matched set. For the second matched set, we impose that the matched male author has the same number of coauthors instead of the same career length as the female author. For the third matched set, we impose that the matched male author has both the same career length and the same number of coauthors.
In these matching experiments, we do not use the authors to whom a research discipline has not been assigned.
Then, we measure the gender imbalance within each set.

We have found that the gender gaps in both total productivity and total impact have drastically decreased after controlling for either the career length or the number of coauthors in all the country groups, except for total impact for China and South Korea (see Fig.~\ref{fig:4}).
When we only use the papers in which the author is in the prominent position,
the gender gaps have considerably decreased in the matching experiments in all the country groups (see Supplementary Section S7 for details).
Therefore, consistent with the previous results \cite{huang2020, li2022}, both the career length and the number of coauthors seem to be key contributors to the gender imbalance in the total productivity and total impact of authors in East Asia as well as in other countries.
Moreover, the gender gaps in total productivity and total impact little decrease after controlling for both the career length and the number of coauthors compared with controlling for just one of them.
These results suggest that the author's career length and number of coauthors are correlated.
In fact, the Pearson correlation coefficient between the career length and the number of coauthors of the authors is roughly 0.4 for each country group (see Supplementary Section S8 for details).

\subsection{Gender imbalance in citation practice}

The gender imbalance in research performance may owe not only to differences between females and males in their research career, such as the career length and the number of coauthors, but also those in citation practices \cite{ferber1988, king2017, kristina2022}.
Therefore, we compare gender imbalance in citation practice between authors in each country group in this section.
We use the 27,616,941 papers each of which falls into four gender categories (i.e., MM, MW, WM, or WW) and has the first and last authors in the same country group (i.e., China, Japan, South Korea, or the other countries). 
Note that we regard sole-author papers by males and females as MM and WW papers, respectively.
There are 66,569 papers (i.e., 53,918 MM, 2,451 MW, 3,945 WM, and 6,255 WW papers) available for this analysis from China, 1,741,061 papers (i.e., 1,547,977 MM, 38,199 MW, 130,235 WM, and 24,650 WW papers) from Japan, 248,388 papers (i.e., 212,509 MM, 6,507 MW, 23,999 WM, and 5,373 WW papers) from South Korea, and 25,560,923 papers (i.e., 16,240,956 MM, 2,000,691 MW, 3,546,901 WM, and 3,772,375 WW papers) from the other countries.

\begin{figure}
  \begin{center}
	\includegraphics[scale=0.17]{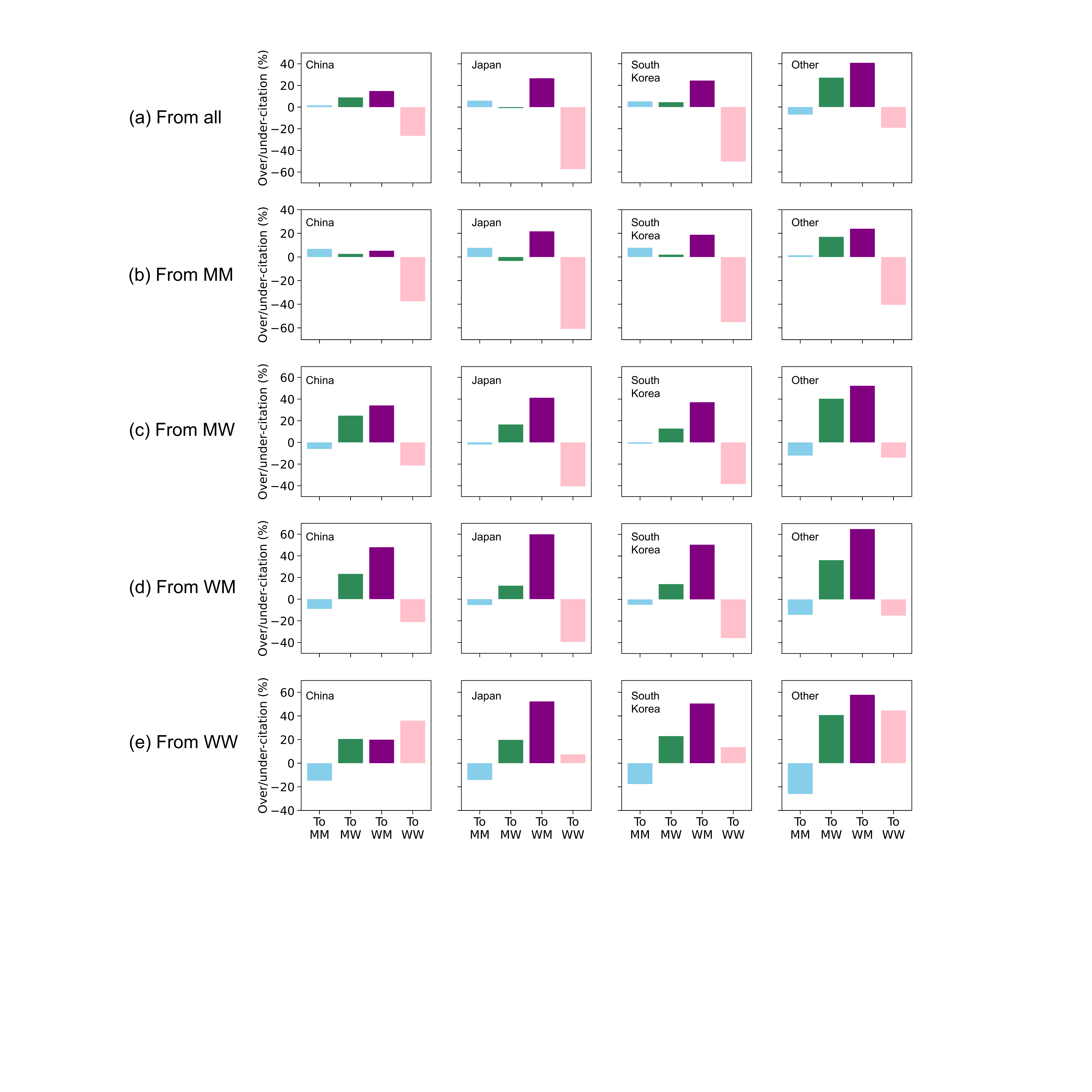}
  \end{center}
  \caption{Gender imbalance in citations made by papers whose authors are in China, Japan, South Korea, and the other countries. (a) All papers. (b) MM papers. (c) MW papers. (d) WM papers. (e) WW papers.}
  \label{fig:5}
\end{figure}

We now compare gender imbalance in citations made by authors in China, Japan, South Korea, and the other countries.
We first count the number of citations made by any papers from each country group to any papers in each gender category (i.e., MM, MW, WM, and WW).
Then, we compare the obtained citation counts with the expected numbers for the random-draws model, which assumes that each paper cites other papers uniformly at random.
We have found that main patterns of over/under-citation of papers in each gender category is qualitatively the same in China, Japan, and South Korea;
papers published in these countries over-cite MM and WM papers and under-cite WW papers (see Fig.~\ref{fig:5}(a)). 
The degree of these over/under-citations is weaker in China than in Japan and South Korea, and slightly larger in Japan than in South Korea (see Fig.~\ref{fig:5}(a)).
Compared with the three East Asian countries, the papers published in the other countries cite less MM papers and more MW, WM, and WW papers (see Fig.~\ref{fig:5}(a)).
Our results of the over-citation of MM papers, which is observed for China, Japan, and South Korea, and the under-citation of WW papers, which is observed for any country group, are consistent with those in previous studies that examined papers from different countries altogether \cite{dworkin2020, fulvio2021, wang2021}.

The amount of over/under-citation made by the papers in each gender category is qualitatively similar among China, Japan, South Korea, and the other countries, which we summarize as follows.
First, the MM papers over-cite MM and WM papers and under-cite WW papers (see Fig.~\ref{fig:5}(b)).
Second, the MW papers under-cite MM and WW papers and over-cite MW and WM papers (see Fig.~\ref{fig:5}(c)).
The WM papers from each country group also show qualitatively the same citation practice (see Fig.~\ref{fig:5}(d)).
Therefore, the MM, MW, and WM papers from each country group under-cite WW papers on average.
Third, in contrast, the WW papers under-cite MM papers and over-cite MW, WM, and WW papers (see Fig.~\ref{fig:5}(e)).
These results are consistent with the previous results for papers published from any countries on neuroscience in that MM, MW, and WM papers under-cite WW papers and that WW papers under-cite MM papers and over-cite WW papers \cite{dworkin2020, wang2021}.
Fourth, papers in any gender category of author tend to cite WM papers more than MW papers (see Figs.~\ref{fig:5}(b)--(e)). 
In sum, we observe consistent gender imbalance with which papers involving males in a prominent position tend to over-cite male papers, broadly defined, 
and papers in which the first and last authors are both females tend to over-cite female papers.

Figure~\ref{fig:5} also reveals differences in over/under-citation behavior among the four country groups. 
First, in any gender category of author, papers from the three East Asian countries under-cite WW papers and over-cite MM papers more strongly than those in the other countries, except for the MM papers from China. Second, papers from the three East Asian countries under-cite MW and WM papers more strongly than in those in the other countries, while we do not have active interpretation of this result.
Third, among the three East Asian countries, in any gender category of author, Japanese and South Korean papers under-cite WW papers more strongly than Chinese papers.

Citation practice is often subject to country bias. For example, health professionals in the US and UK tend to over-cite articles published in a medical journal of their own countries \cite{campbell1990}; in nanotechnology, Chinese authors are more likely to cite each other than US authors do \cite{tang2015}.
Motivated by these previous findings, we hypothesize that papers from China, Japan, and South Korea cite papers from the same country in a different gender-biased manner compared to when citing papers from the other countries.
To investigate this possibility, we decompose citations made by the papers from each country into domestic citations (i.e., citations to the papers published in the same country as the authors') and foreign citations (i.e., citations to the papers published in different countries), and calculate gender imbalance separately for domestic and foreign citations for each country.

It turns out that patterns of the over/under-citations for the foreign citations are qualitatively the same as those for the overall citations (see Fig.~\ref{fig:5} and Supplementary Fig.~S3). 
This result is expected because the foreign citations are yet dominant in China, Japan, and South Korea; they account for 98.2\% of the overall citations in China, 77.8\% in Japan, and 93.4\% in South Korea.
We now turn to domestic citations made by the papers from China, Japan, and South Korea.
First, while the papers from China under-cite WW papers on average
(see Fig.~\ref{fig:5}(a)), they over-cite WW papers from China (see Supplementary Fig.~S2(a)).
Second, papers in any gender category and any country are more likely to cite WW papers from the same country than they cite WW papers from any country, except for MM papers from Japan (see Figs.~\ref{fig:5}(b)--(e) and Supplementary Figs.~S2(b)--(e)).
Third, WW papers from each country are more likely to cite papers involving female authors in a prominent position (i.e., MW, WM, or WW papers) in the same country than they cite those in any country (see Fig.~\ref{fig:5}(e) and Supplementary Fig.~S2(e)).
In fact, on average, the WW papers from China, Japan, and South Korea over-cite the MW, WM, or WW papers from the same country at least twice more than they over-cite the MW, WM, or WW papers in general.
In particular, the WW papers from Japan and South Korea over-cite the WW papers from the same country 56.8 and 35.8 times, respectively, more than they over-cite WW papers in general.
These results are also consistent with the previous findings that authors tend to cite papers from the same country \cite{campbell1990, bakare2017, baccini2019}
and that the author's gender and another variable (e.g., authors' race and country) can have combined effects on gender imbalance in citation behavior~\cite{kozlowski2022, zheng2023}.

\section{Discussion} \label{section:4}

In this study, we hypothesized that China, Japan, and South Korea are exposed to larger gender imbalance in academia than other countries.
Therefore, we compared gender imbalance among authors in the four country groups (i.e., China, Japan, South Korea, and the other countries).
We found that the fractions of female authors in China, Japan, and South Korea are smaller than that in the other countries.
We also found that the gender imbalance in research career is notably larger in Japan and is present but weaker in China and South Korea than in the other countries.
In terms of citation practice, papers from the three East Asian countries are more likely to under-cite WW papers (i.e., papers in which women occupy the prominent author positions) than those in the other countries do.
These results are largely consistent with the previous results in that Asian countries tend to be exposed to larger gender imbalance in academia than other countries  \cite{lariviere2013, holman2018, bendels2018, thelwall2019, chan2020, gender_report_2020, meho2022, el2022}.
Because East Asia accounts for a substantial fraction of research expenditure, number of researchers, and research output in the world, gender imbalance in academia in East Asia is considered to affect our understanding of the global landscape of gender imbalance in academia.

We also hypothesized that Japan is exposed to larger gender imbalance in academia than China and South Korea.
We compared the gender imbalance among authors in China, Japan, and South Korea aided by a unified and high-accuracy gender assignment method.
In our data set, the fraction of female authors was the largest in China and smallest in South Korea; in previous studies, it was the largest in China \cite{lariviere2013, holman2018, chan2020, meho2022} and the smallest in Japan \cite{lariviere2013, holman2018} or South Korea \cite{chan2020, meho2022}.
Therefore, our results are roughly consistent with theirs.
We also found that the gender imbalance in research career and the degree of under-citation of any WW papers are larger in Japan than in China and South Korea.
Haghani et al.~suggested that gender imbalance in total productivity has been larger in Japan than in South Korea between 2006 and 2020 \cite{haghani2022}, which is consistent with our results. Due to a poor accuracy at identifying the gender from East Asian names, other studies either failed to sample a sufficient number of gender-assigned researchers in the three East Asian countries or excluded those researchers in an early stage of their analysis \cite{bendels2018, gender_report_2020, huang2020}. 
In contrast, we addressed this technical barrier to comprehensively compare gender imbalance in the academia of three East Asian countries in terms of the number of researchers, research career, and citation practice of individual researchers. Our results are a significant addition to the previous studies that examined the three countries but only in terms of the number of researchers.

For bibliographic data analysis involving gender comparison, one usually needs to identify authors' gender from their names \cite{lariviere2013, west2013, huang2020}.
This task is difficult for East Asian names \cite{bendels2018, santamaria2018, jadidi2018}. 
We built an algorithm to detect the gender of Chinese, Japanese, and Korean names with a classification accuracy of at least 90\% on our benchmark data. In the course of parameter tuning, we found that, given the parameter values, the classification accuracy depends on the era, which is quantified by the year of the first publication, as well as the country. For example, for the authors in China, we needed to impose a higher threshold of `accuracy', $\theta$, when the names of more recent than old authors are input. This phenomenon implies that the gender of more recent Chinese names is more difficult to detect. In contrast, it was more difficult to detect the gender of older than more recent authors in Japan.
We also observed that tuning the threshold on the number of samples in the database, $n_s$, can improve the accuracy of gender inference for East Asian names; other studies used the same technique before for author names from various countries~\cite{thelwall2020, meho2022}.
Despite our efforts, the fraction of gender-assigned authors in China (i.e., 5.0\%) was much lower than that in Japan and South Korea (i.e., 60.3\% and 43.7\%, respectively).
Therefore, our Chinese data may not be representative of the entire population of Chinese researchers.
It is notably difficult to infer gender from Chinese names written in English \cite{karimi2016, bendels2018, gender_report_2020}.
In fact, previous studies failed to collect sufficiently many samples of Chinese gender-assigned authors when their names were written in English \cite{holman2018, bendels2018}. Another study used 31,150 samples of Chinese gender-assigned authors whose names were written in Mandarin \cite{gender_report_2020}. The fraction of the gender-assigned authors in the latter study was 40.7\% ($\approx$ 31,150/76,627), which is much larger than the fraction for our Chinese samples. Similarly,
case studies in Italy \cite{abramo2021}, Norway \cite{abramo2021}, and Poland \cite{kwiek2021} exploited the researcher's gender provided by local academic databases to enhance the accuracy of gender assignment.
Given these motivating results, investigating gender imbalance in Chinese academia using data with names and other information in the Chinese language warrants future work. 

We did not systematically examine different gender inference tools.
We employed Gender API, which yielded the highest accuracy on average for Asian names on a benchmark data set \cite{santamaria2018}.
In fact, there are more recent developments in gender inference tools.
For example, Van Buskirk et al.\,developed an open-source gender inference tool built on publicly available data spanning over 150 countries and more than a century \cite{vanbuskirk2023}.
Antonoyiannakis et al.\,proposed a gender inference method to address the issue of overestimating the gender likelihood for rare names by relaxing a conventional sampling assumption on names \cite{antonoyiannakis2023}.
Combining our gender-inference algorithm with such tools may improve the accuracy of gender inference for East Asian names.
However, we should also note that the accuracy of gender inference tools is approaching a plateau \cite{vanbuskirk2023} and that error rates are heterogeneously distributed for names with different languages, races, and ethnicities \cite{lockhart2023}, which both pose challenges.

Recent studies suggested that the strength of gender imbalance varies by the academic author's race and ethnicity \cite{bertolero2020, kozlowski2022}.
Similar to these studies, we found that the strength of gender imbalance is partially attributed to the author's nationality (i.e., the author's country of origin inferred from their first name) in the three East Asian countries.
Alesina et al.\,introduced ``ethnic inequality'' measured by within-country differences in well-being across ethnic groups and compared it across 173 countries including the three East Asian countries \cite{alesina2016}.
Their results suggested that ethnic inequality is much larger in China than in Japan and South Korea \cite{alesina2016}.
It may be fruitful to measure ethnicity inequality in academia and investigate interaction between gender imbalance and ethnicity inequality in academia in East Asia and elsewhere. 
On the other hand, inferring an author's race, ethnicity, or nationality may incur a bias. In fact, a recent study found that thresholding-based approaches to inference of races from the author's name are not accurate \cite{kozlowski2022_2}.
The accuracy of inferring the author's nationality may be similarly compromised for some nations in our samples. Detailed analyses of biases in the inference of author's nationality warrant future work.

There remains a long-standing need to collect and analyze data on gender representation across different aspects of academia and to evaluate policy interventions aiming at improving gender imbalance \cite{llorens2021}.
We limited our analyses to gender imbalance in academia in terms of the number of researchers, research careers, and citation practices, among many issues and indices.
Other issues include gender diversity in research teams \cite{macaluso2016, yang2022, ross2022, zheng2022}, research careers of female researchers who have children \cite{morgan2021, zheng2022_2, esther2023}, associations between gender, race, and ethnicity imbalances in academia \cite{bertolero2020, kozlowski2022}, which we discussed above, underrepresentation of female researchers in peer-review processes \cite{fox2016, helmer2017, murray2019, squazzoni2021}, and gender imbalance in terms of the effects of the COVID pandemic around 2020 on research careers \cite{myers2020, minello2021, deryugina2021}.

Quantitative tracking tools for addressing gender imbalance in academia have also been proposed \cite{llorens2021}.
Examples include recent efforts to address gender imbalance in citation practices of individual authors using a so-called Citation Diversity Statement and cleanBib tool \cite{zhou2020, zurn2022}. 
The Gender Balance Assessment Tool is a tool to quantify gender imbalance in words appearing in documents (e.g., a conference program) \cite{sumner2018}.
NamePrism is a tool to infer the ethnicity, nationality, or race from a given name \cite{ye2017}.
Such tools are expected to promote our quantitative understanding of gender imbalance in academia.
For example, Ruzycki et al.\,used the tool proposed in Ref.~\cite{sumner2018} to quantify gender imbalance in speakers at major academic medical conferences held in Canada and the United States between 2007 and 2017 \cite{ruzycki2019}. 
Webber et al.\,also used the same tool to quantify gender bias in words within documents on evaluations of pediatric faculty \cite{webber2022}.
Liu et al.\,used NamePrism to quantify race imbalance in editorial board members of journals published by six publishers \cite{liu2023_2}.
Deploying these and similar tracking tools to investigating regional differences in gender and race imbalance in academia may be fruitful.

Our study has additional limitations. First, when analyzing the research career, we focused on the authors whose last publication year was 2015 or earlier.
Therefore, our data set may not reflect recent efforts to support the academic participation of women in East Asian countries (e.g., China \cite{china_nature_comment}, Japan \cite{japan_nature_news}, and South Korea \cite{korea_the_news}).
Moreover, recent changes in academic environments in China are remarkable, as evidenced by the rapid growth of research outputs \cite{zhou2006, xie2014, gomez2022}, the number of doctoral students \cite{cao2019}, and the number of researchers who returned to China after receiving academic training in foreign countries \cite{cao2019}.
Because we needed to discard many researchers in China because of the difficulty in gender identification, further work is needed to verify our findings on China.
Second, we only used the OpenAlex database. Because OpenAlex is a successor of MAG, our results may be biased due to the problems that the MAG had. 
For example, MAG database covers a larger number of publications but misses more citations than other data (e.g., Scopus and Web of Science) do~\cite{visser2021}.
Third, our analysis only focused on researchers writing academic papers.
A recent study suggested that Japanese children acquire the gender stereotype to associate brilliance with males after they start to go to school \cite{okanda2022}. 
It is also important to investigate gender imbalance in wider education than has been investigated in the present study (e.g., Ref.~\cite{evans2021}).
Fourth, statistical modeling including that for comparing the results among countries may help us to identify further features of gender imbalance in East Asian academia.
Addressing these and other limitations is expected to help us better quantify and improve gender imbalance in academic systems and in practices of individual researchers.

\section*{Acknowledgments}

Kazuki Nakajima was supported in part by JSPS KAKENHI Grant Number JP21J10415. Naoki Masuda was supported in part by the Japan Science and Technology Agency (JST) Moonshot R\&D (under Grant Number JPMJMS2021) and in part by JSPS KAKENHI under Grant Numbers JP21H04595 and JP23H03414.

\section*{Declaration of Competing Interest}
The authors declare no competing interests.

\newpage

\begin{center}
\vspace*{12pt}
{\Large Supplementary Materials for:\\
\vspace{12pt}
Quantifying gender imbalance in East Asian academia: Research career and citation practice}
\vspace{12pt} \\
\end{center}

\setcounter{figure}{0}
\setcounter{table}{0}
\setcounter{section}{0}

\renewcommand{\thesection}{S\arabic{section}}
\renewcommand{\thefigure}{S\arabic{figure}}
\renewcommand{\thetable}{S\arabic{table}}

\begin{center}
\author{Kazuki Nakajima, Ruodan Liu, Kazuyuki Shudo, and Naoki Masuda}
\vspace{24pt} \\
\end{center}

\section{Research discipline of papers and authors}

Each paper in our data set is associated with at least one concept.
There are 65,073 concepts in the data set, which form a six-level tree structure from level 0 to 5.
Lower-level concepts are more general, and higher-level concepts are more specific.
For example, if we traverse downward the tree structure from the ``computer science'' concept at level 0, we find a level-5 concept named ``Java bytecode''.
There are 19 concepts at level 0 (i.e., art, biology, business, chemistry, computer science, economics, engineering, environmental science, geography, geology, history, materials science, mathematics, medicine, philosophy, physics, political science, psychology, and sociology). 
We use the level-0 concepts as research disciplines.

We assigned one research discipline to each paper $z$ as follows.
We first count the frequency of each discipline by fully traversing the tree upward from each of the $z$'s concepts to level 0.
If the most frequent discipline is unique, we assign it to $z$. Otherwise, we exclude $z$ from the data set.

Then, we assigned one research discipline to each author $u$ based on $u$'s papers.
We first count the frequency of each research discipline across all papers authored by $u$.
If the most frequent discipline is unique, we assign it to $u$. Otherwise, we do not assign any discipline to $u$.

\section{Country with which the author is affiliated}

We assigned a single country to each author $u$ based on $u$'s affiliation as follows.
The country code (i.e., the ISO 3166-1 alpha-2 country code) is available for most affiliations (i.e., 99.1\%) present in our data set.
Therefore, we first counted the frequency with which each country appears in $u$'s affiliation over all papers authored by $u$.
Note that $u$'s affiliation data may be missing for some papers authored by $u$.
Then, if the most frequent country is unique, we assign it to $u$. Otherwise, we exclude $u$ from the data set.

\section{Assessing the accuracy of gender assignment}

\subsection{Six baseline methods and classification accuracy}

In response to a first name given as input, Gender API returns either `female', `male', or `unknown', its `accuracy' (which is a terminology of Gender API and different from the classification accuracy on the test set discussed below; therefore we put the quotation marks here and in the following text), and the number of samples of the given first name in the database. The country name is an optional input to Gender API.
Our gender assignment method using Gender API consists of (i) whether or not we input the author's country name, (ii) the threshold on the `accuracy', denoted by $\theta$, such that the API's outputs with an `accuracy' less than $\theta$ are discarded, and (iii) the threshold on the number of samples, denoted by $n_s$, such that
the API's outputs with less than $n_s$ samples are discarded.

To assess the classification accuracy of a gender assignment method, we manually checked the correctness of the assigned gender of sampled authors as follows.
We first classified the authors that have at least one candidate of the first name into four groups, i.e., China, Japan, South Korea, and the other countries, according to their country.
Second, we classified the authors in each country group into the following four subgroups according to the year of their first publication: (i) 1990 or before, (ii) between 1991 and 2000, (iii) between 2001 and 2010, and (iv) between 2011 and 2020.
Third, we selected the top 25 authors in terms of the number of published papers for each gender, each of the four country groups, and each of the four year subgroups, which yielded 100 females and 100 males for each country group.
Then, we searched their publicly available electronic materials (e.g., curriculum vitae, profiles on social networking services, and web pages with face photos) based on their name and affiliation to manually label each of these authors either ``female'', ``male'', and ``N/A''. 
The second author (i.e., RL), who is native speaker of Chinese, carried out this task for China.
The first author (i.e., KN), who is native speaker of Japanese, carried out this task for the other three country groups.
Their gender assignment decision is admittedly subjective.
Label ``N/A'' indicates that we could not find sufficient online materials for the author to judge their gender.

Our goal is to find a gender assignment method that yields a classification accuracy of at least 90\% for each gender, each country group, and each year subgroup.
For each country group, we first examined six baseline methods: ``Global'', ``Local'', ``Hybrid'', ``Global$\alpha$'', ``Local$\alpha$'', and ``Hybrid$\alpha$''.
In the Global method, for each author, we feed each candidate of their first name to the API without specifying their country.
Then, among the API's outputs, we kept only the outputs for which the `accuracy' that the API returns is at least $\theta$ and the number of samples is at least $n_s$.
We set $\theta=$ 90\% and $n_s= 1$.
We then find the largest `accuracy' value among all the input first-name candidates that return female as output.
Similarly, we find the largest `accuracy' value among all the input first-name candidates that return male as output.
If the former is larger than the latter, we assign female to the author.
If the former is smaller than the latter, we assign male to the author.
If the former is equal to the latter, then we do not assign any gender to the author.

The only difference between the Global, Local, and Hybrid methods is the inputs to the API.
In the Local method, for each author, we feed each candidate of their first name to the API but not their country.
In the Hybrid method, for each author, we feed each candidate of their first name to the API without specifying their country, and then we again feed the same first name to the API by specifying their country.
Moreover, we define the Global$\alpha$, Local$\alpha$, and Hybrid$\alpha$ methods by changing $n_s = 1$ to $n_s =$ 10 in the Global, Local, and Hybrid methods, respectively.
Raising $n_s$ may improve the classification accuracy; the same technique was used before~\cite{thelwall2020, meho2022}.
The method with the best classification accuracy among the six baseline methods does not reach 90\% classification accuracy for some pairs of the country group and year subgroup. In this case, we examine additional methods in which we separately adjust $\theta$ or $n_s$ just for the pairs with low classification accuracy, as we describe in the remainder of this section, until the classification accuracy reaches 90\% in all the cases.

\subsection{Authors in China}

We found that the Local method yields the highest classification accuracy on average over the eight subgroups (i.e., four year subgroups for each gender) among the Global, Local, and Hybrid methods (see Table \ref{table:s1}). 
The Local method inferred the gender of male authors in each year subgroup with more than 90\% classification accuracy.
However, the classification accuracy for the females with the same method was much lower than 90\%.
Then, we additionally considered Global$\alpha$, Local$\alpha$, and Hybrid$\alpha$.
These variants improve the classification accuracy, and Local$\alpha$ yields the highest average of the classification accuracy over the eight subgroups (see Table \ref{table:s1}).
However, the classification accuracy of Local$\alpha$ is still much lower than 90\% for the females whose first publication year is 1991 or later.

Therefore, we considered three additional variants of Local$\alpha$, named Local$\beta$, Local$\gamma$, and Local$\delta$.
Local$\beta$ is the same as Local$\alpha$ except that, in Local$\beta$, we set $\theta=$ 95\% for the authors whose first publication year is 1991 or later.
Then, the classification accuracy for the females whose first publication year is 2011 or later is higher than 90\%; however, that for the females whose first publication year is between 1991 and 2010 is not (see Table \ref{table:s1}).
Next, we considered Local$\gamma$, which is the same as Local$\beta$ except that Local$\gamma$ uses $\theta=$ 99\% for the authors whose first publication year is between 1991 and 2010.
With Local$\gamma$, the classification accuracy for the females whose first publication year is between 1991 and 2000 is higher than 90\%; however, that for the females whose first publication year is between 2001 and 2010 is not (see Table \ref{table:s1}).
Therefore, we considered Local$\delta$, which is the same as Local$\gamma$ except that Local$\delta$ uses $n_s = 50$ for the females whose first publication year is between 2001 and 2010.
Then, the classification accuracy for the females whose first publication year is between 2001 and 2010 is higher than 90\% (see Table \ref{table:s1}).
Therefore, we used Local$\delta$, which yields at least 90\% classification accuracy for the females and males in each year subgroup.

\subsection{Authors in Japan}

For the authors in Japan, Local$\alpha$ yielded the highest average of the classification accuracy over the eight subgroups among the six baseline methods.
Local$\alpha$ detected the gender of the males in each year subgroup and of the females whose first publication year is 1991 or later with more than 90\% accuracy (see Table \ref{table:s2}). However, the classification accuracy for the females whose first publication year is 1990 or before is much smaller than 90\% with Local$\alpha$.

Therefore, we considered two additional variants of Local$\alpha$, named Local$\epsilon$ and Local$\zeta$.
In Local$\epsilon$, we changed Local$\alpha$ by setting $\theta=$ 95\% just for the authors whose first publication year is 1990 or before.
We found that, with Local$\epsilon$, the classification accuracy for the females whose first publication year is 1990 or before is still lower than 90\% (see Table \ref{table:s2}).
In Local$\zeta$, we changed Local$\epsilon$ by setting $\theta=$ 99\% just for the authors whose first publication year is 1990 or before.
The classification accuracy with Local$\zeta$ is higher than 90\% for the females whose first publication year is 1990 or before (see Table \ref{table:s2}).
Therefore, we used Local$\zeta$, which yields at least 90\% classification accuracy for the females and males in each year subgroup.

\subsection{Authors in South Korea}

For the authors in South Korea, Local$\alpha$ yielded the highest average of the classification accuracy over the eight subgroups among the six baseline methods, which 
was higher than 90\% for the females and males in each year subgroup (see Table \ref{table:s3}). Therefore, we used Local$\alpha$.

\subsection{Authors in the other countries}

For the authors in the other countries, Global and Global$\alpha$ yielded the highest average of the classification accuracy over the eight subgroups among the six baseline methods.
Both methods yielded an classification accuracy higher than 90\% for the females and males in each subgroup (see Table \ref{table:s4}).
Note that Global$\alpha$ yielded exactly the same classification accuracy for the females and males in each subgroup as Global because all the 200 sampled authors have first names that have more than ten samples in the database. As we did for the authors in China, Japan, and South Korea, and as previous studies also did \cite{thelwall2020, meho2022}, we decided to use Global$\alpha$ for the authors in the other countries.

\section{Citation count and its normalization}

We count and normalize the number of citations that each paper has received using the method proposed in Ref.~\cite{huang2020} as follows.
For each paper $z$, we first count the number of citations received by $z$ within 10 years after its publication without self-citations (i.e., citations to the papers whose authors overlap those of $z$).
Then, we define the citation impact of $z$ as the number of citations divided by the average number of citations received by the papers published in the same year.

\section{Fraction of female authors by research discipline}

Table~\ref{table:s5} shows the number of gender-assigned authors and the fraction of female authors in China, Japan, South Korea, and the other countries by author's research discipline.

\section{Statistical significance of gender gap in research career}

We run the unpaired two-tailed Welch's $t$-test for the combination of an indicator and a set of authors, as was done in Ref.~\cite{huang2020}.
The $t$-test decides whether the female and male authors in the set, which have unequal sample sizes and unequal variances of the indicator, deviate from the null hypothesis that the female and male authors have the same mean of the indicator.
For the female authors in the set, we denote by $n_{\text{F}}$, $\mu_{\text{F}}$, and $s_{\text{F}}$ the number of samples, the sample mean of the indicator, and the corrected sample standard deviation of the indicator, respectively. 
Similarly, we denote the corresponding quantities for the male authors by $n_{\text{M}}$, $\mu_{\text{M}}$, and $s_{\text{M}}$.
Then, we calculate the $t$-statistic given by
\begin{align}
t = \frac{\mu_{\text{F}} - \mu_{\text{M}}}{\sqrt{\frac{s_{\text{F}}^2}{n_{\text{F}}}} +\sqrt{\frac{s_{\text{M}}^2}{n_{\text{M}}}}},
\end{align}
the $p$-value of the obtained $t$-statistic against the null hypothesis, and the 95\% confidence interval (CI) of $\mu_{\text{F}} - \mu_{\text{M}}$.
We use `t.test' function from `stats' package in R \cite{r_manual} to calculate the $t$-statistic, $p$-value, and CI.

Tables~\ref{table:s6}, \ref{table:s7}, \ref{table:s8}, and \ref{table:s9} show the results of the unpaired two-tailed Welch's $t$-test for the authors in China, Japan, South Korea, and the other countries. The $p$-values less than $0.001$ in these tables remain significant after the Bonferroni correction at a significance level of $0.05$ because there are 42 comparisons in total in these four tables.

\section{Matching experiments for the papers in which the authors occupy the prominent position}

\begin{figure}[p]
  \begin{center}
	\includegraphics[scale=0.21]{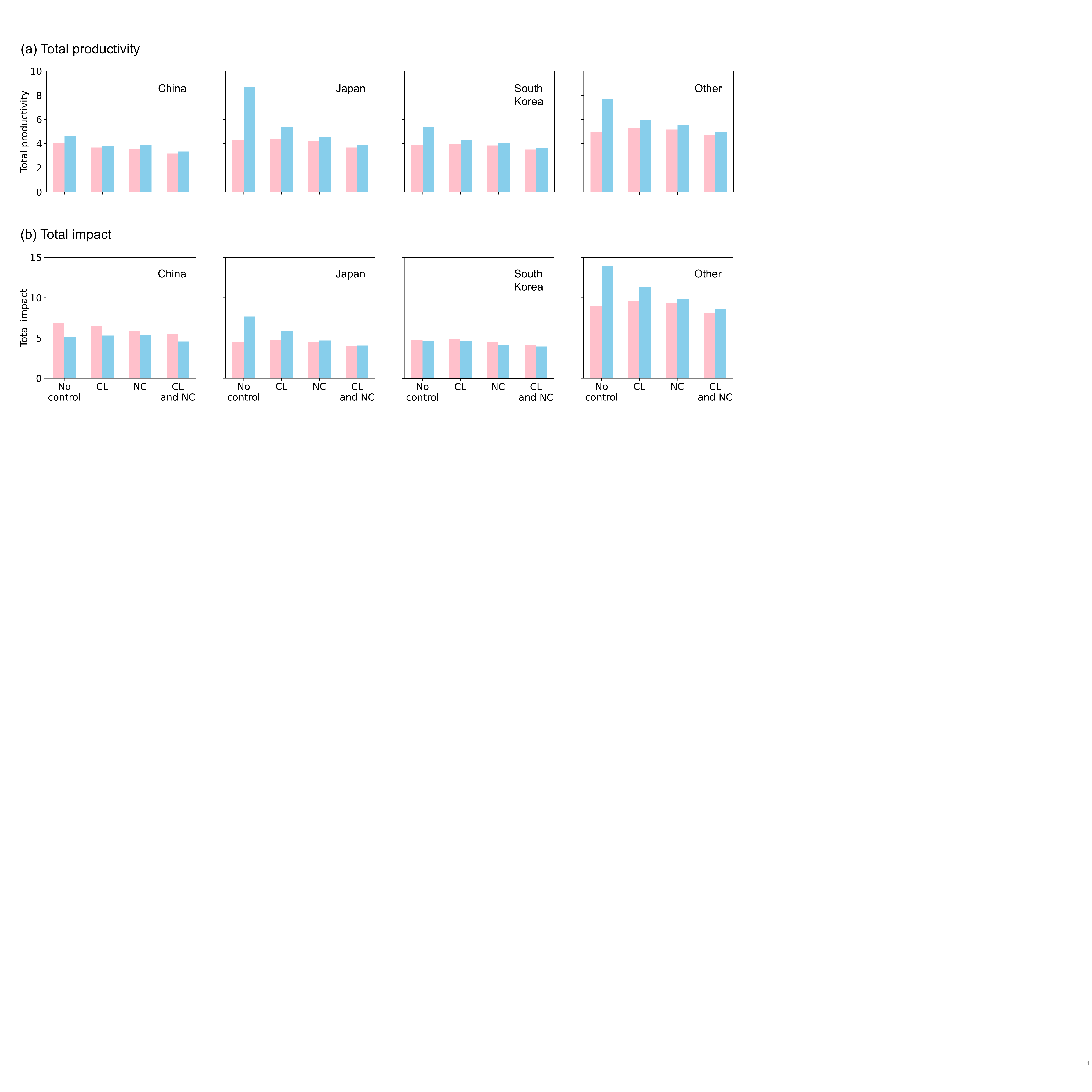}
  \end{center}
  \caption{Results of the matching experiment when we restrict the analysis to the papers in which the authors occupy the prominent author position. (a) Total productivity. (b) Total impact. In the matching experiment, we controlled for the author's career length (CL) and/or their number of coauthors (NC) in addition to their country group, year of the first publication, and research discipline.}
  \label{fig:s1}
\end{figure}

Figure~\ref{fig:s1} shows the results for the matching experiments in which we restrict the analysis to the papers in which the authors occupy the prominent author position.

\section{Correlation between author's career length and number of coauthors}

Table~\ref{table:s10} shows the Pearson correlation coefficients between the career length and the number of coauthors for the set of all the authors, that of female authors, and that of male authors in each country group.

\section{Gender imbalance in domestic and foreign citations}

\begin{figure}[p]
  \begin{center}
	\includegraphics[scale=0.2]{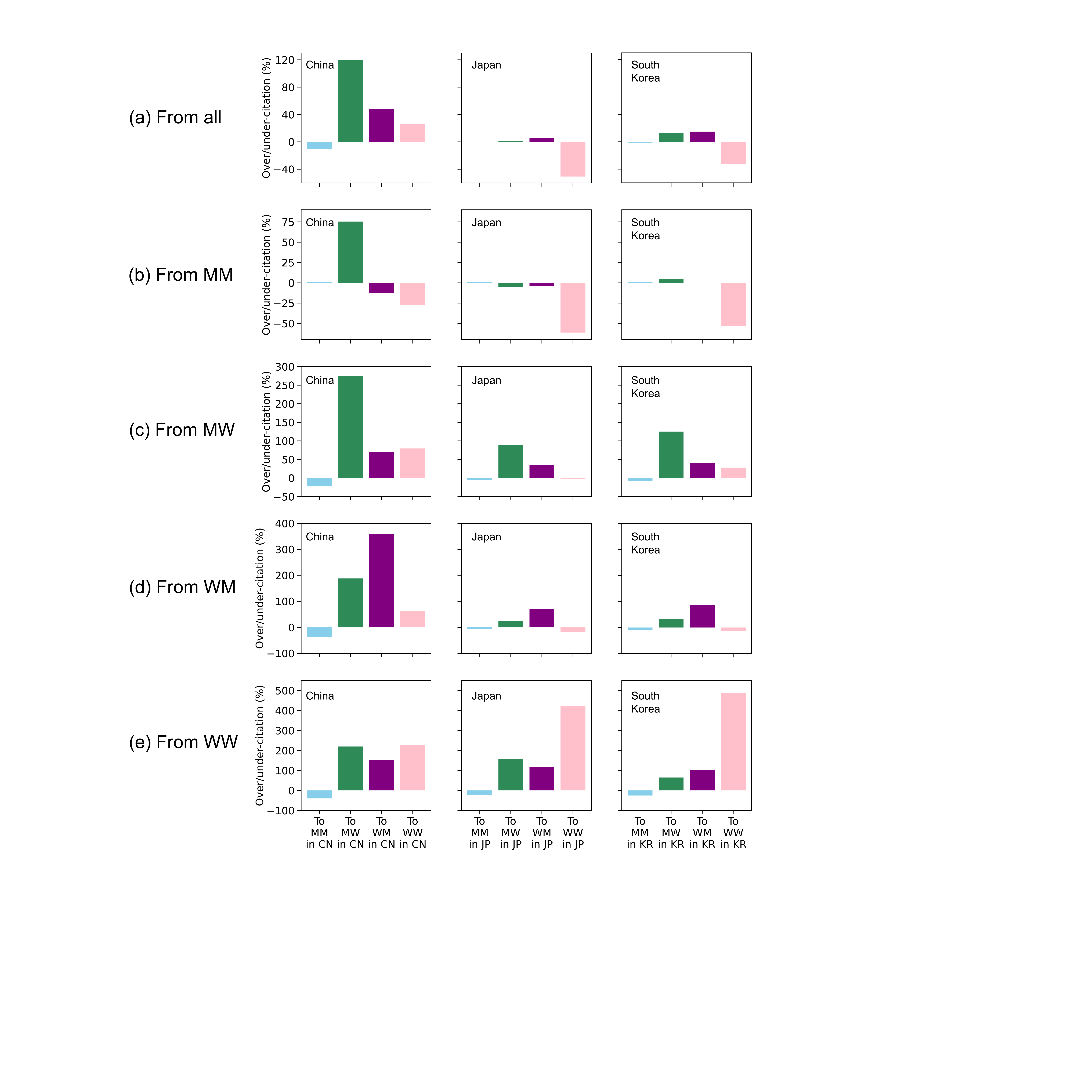}
  \end{center}
  \caption{Gender imbalance in domestic citations made by papers whose authors are in China, Japan, and South Korea. (a) All papers. (b) MM papers. (c) MW papers. (d) WM papers. (e) WW papers. CN, JP, and KR stand for China, Japan, and South Korea, respectively.}
  \label{fig:s2}
\end{figure}

\begin{figure}[p]
  \begin{center}
	\includegraphics[scale=0.2]{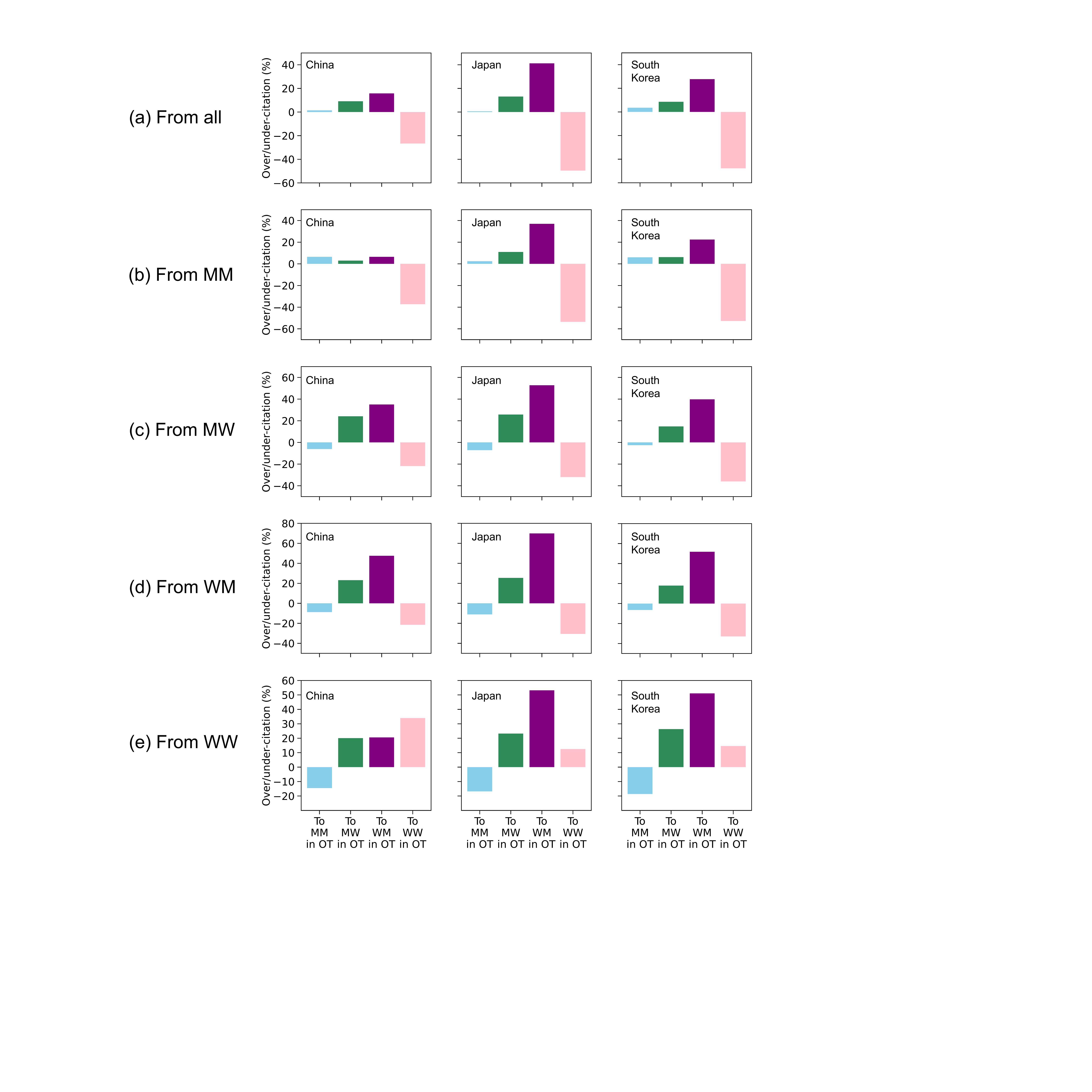}
  \end{center}
  \caption{Gender imbalance in foreign citations made by papers whose authors are in China, Japan, and South Korea. (a) All papers. (b) MM papers. (c) MW papers. (d) WM papers. (e) WW papers. OT stands for the other countries.}
  \label{fig:s3}
\end{figure}

Figure~\ref{fig:s2} shows the over/under-citations for domestic citations made by papers from China, Japan, and South Korea.
Figure~\ref{fig:s3} shows the corresponding results for foreign citations.

\newpage

\begin{table}[p]
\caption{Classification accuracy of the different gender assignment methods for the authors in China. In each cell, we report the classification accuracy for female and male authors in red and blue, respectively. The number after the classification accuracy shows the number of authors among the 25 selected authors for whom we were able to manually assign the binary gender.}
\label{table:s1}
\begin{center}
\begin{tabular}{| D | E | E | E | E |} \hline
\multirow{2}{*}{Method} & \multicolumn{4}{c|}{Year of the first publication} \\ \cline{2-5}
& 1990 or before & 1991--2000 & 2001--2010 & 2011--2020 \\ \hline
\multirow{2}{*}{Global} & \textcolor{red}{64.0\%} (25) & \textcolor{red}{48.0\%} (25) & \textcolor{red}{54.1\%} (24) & \textcolor{red}{73.9\%} (23) \\
& \textcolor{blue}{95.8\%} (24) & \textcolor{blue}{96.0\%} (25) & \textcolor{blue}{100\%} (25) & \textcolor{blue}{100\%} (24) \\
\multirow{2}{*}{Local} & \textcolor{red}{72.0\%} (25) & \textcolor{red}{60.0\%} (25) & \textcolor{red}{65.2\%} (23) & \textcolor{red}{84.0\%} (25) \\
& \textcolor{blue}{100\%} (21) & \textcolor{blue}{96.0\%} (25) & \textcolor{blue}{100\%} (25) & \textcolor{blue}{100\%} (25) \\
\multirow{2}{*}{Hybrid} & \textcolor{red}{68.0\%} (25) & \textcolor{red}{36.0\%} (25) & \textcolor{red}{56.5\%} (23) & \textcolor{red}{68.0\%} (25) \\
& \textcolor{blue}{100\%} (22) & \textcolor{blue}{96.0\%} (25) & \textcolor{blue}{100\%} (25) & \textcolor{blue}{100\%} (25) \\
\hline
\multirow{2}{*}{Global$\alpha$} & \textcolor{red}{84.0\%} (25) & \textcolor{red}{88.0\%} (25) & \textcolor{red}{66.6\%} (24) & \textcolor{red}{69.5\%} (23) \\
& \textcolor{blue}{95.6\%} (23) & \textcolor{blue}{96.0\%} (25) & \textcolor{blue}{100\%} (25) & \textcolor{blue}{100\%} (24) \\
\multirow{2}{*}{Local$\alpha$} & \textcolor{red}{92.0\%} (25) & \textcolor{red}{88.0\%} (25) & \textcolor{red}{78.2\%} (23) & \textcolor{red}{87.5\%} (24) \\
& \textcolor{blue}{100\%} (21) & \textcolor{blue}{96.0\%} (25) & \textcolor{blue}{100\%} (25) & \textcolor{blue}{100\%} (25) \\
\multirow{2}{*}{Hybrid$\alpha$} & \textcolor{red}{80.0\%} (25) & \textcolor{red}{88.0\%} (25) & \textcolor{red}{73.9\%} (23) & \textcolor{red}{72.0\%} (25) \\
& \textcolor{blue}{100\%} (22) & \textcolor{blue}{96.0\%} (25) & \textcolor{blue}{100\%} (25) & \textcolor{blue}{100\%} (24) \\
\hline
\multirow{2}{*}{Local$\beta$} & \textcolor{red}{92.0\%} (25) & \textcolor{red}{84.0\%} (25) & \textcolor{red}{82.6\%} (23) & \textcolor{red}{95.8\%} (24) \\
& \textcolor{blue}{100\%} (21) & \textcolor{blue}{96.0\%} (25) & \textcolor{blue}{100\%} (25) & \textcolor{blue}{100\%} (25) \\
\multirow{2}{*}{Local$\gamma$} & \textcolor{red}{92.0\%} (25) & \textcolor{red}{92.0\%} (25) & \textcolor{red}{83.3\%} (24) & \textcolor{red}{95.8\%} (24) \\
& \textcolor{blue}{100\%} (21) & \textcolor{blue}{96.0\%} (25) & \textcolor{blue}{95.8\%} (24) & \textcolor{blue}{100\%} (25) \\
\multirow{2}{*}{Local$\delta$} & \textcolor{red}{92.0\%} (25) & \textcolor{red}{92.0\%} (25) & \textcolor{red}{96.0\%} (25) & \textcolor{red}{95.8\%} (24) \\
& \textcolor{blue}{100\%} (21) & \textcolor{blue}{96.0\%} (25) & \textcolor{blue}{95.8\%} (24) & \textcolor{blue}{100\%} (25) \\
\hline
\end{tabular}
\end{center}
\end{table}

\begin{table}[p]
\caption{Classification accuracy of the different gender assignment methods for the authors in Japan.}
\label{table:s2}
\begin{center}
\begin{tabular}{| D | E | E | E | E |} \hline
\multirow{2}{*}{Method} & \multicolumn{4}{c|}{Year of the first publication} \\ \cline{2-5}
& 1990 or before & 1991--2000 & 2001--2010 & 2011--2020 \\ \hline
\multirow{2}{*}{Global} & \textcolor{red}{48.0\%} (25) & \textcolor{red}{91.6\%} (24) & \textcolor{red}{87.5\%} (24) & \textcolor{red}{90.4\%} (21) \\
& \textcolor{blue}{100\%} (25) & \textcolor{blue}{100\%} (24) & \textcolor{blue}{100\%} (22) & \textcolor{blue}{100\%} (25) \\
\multirow{2}{*}{Local} & \textcolor{red}{56.0\%} (25) & \textcolor{red}{91.6\%} (24) & \textcolor{red}{91.6\%} (24) & \textcolor{red}{91.6\%} (24) \\
& \textcolor{blue}{100\%} (25) & \textcolor{blue}{100\%} (24) & \textcolor{blue}{100\%} (22) & \textcolor{blue}{100\%} (24) \\
\multirow{2}{*}{Hybrid} & \textcolor{red}{48.0\%} (25) & \textcolor{red}{91.6\%} (24) & \textcolor{red}{83.3\%} (24) & \textcolor{red}{87.5\%} (24) \\
& \textcolor{blue}{100\%} (25) & \textcolor{blue}{100\%} (24) & \textcolor{blue}{100\%} (22) & \textcolor{blue}{100\%} (24) \\
\hline
\multirow{2}{*}{Global$\alpha$} & \textcolor{red}{56.0\%} (25) & \textcolor{red}{91.6\%} (24) & \textcolor{red}{87.5\%} (24) & \textcolor{red}{95.2\%} (21) \\
& \textcolor{blue}{100\%} (25) & \textcolor{blue}{100\%} (24) & \textcolor{blue}{100\%} (22) & \textcolor{blue}{100\%} (25) \\
\multirow{2}{*}{Local$\alpha$} & \textcolor{red}{64.0\%} (25) & \textcolor{red}{91.6\%} (24) & \textcolor{red}{95.8\%} (24) & \textcolor{red}{91.3\%} (23) \\
& \textcolor{blue}{100\%} (25) & \textcolor{blue}{100\%} (24) & \textcolor{blue}{100\%} (22) & \textcolor{blue}{100\%} (24) \\
\multirow{2}{*}{Hybrid$\alpha$} & \textcolor{red}{56.0\%} (25) & \textcolor{red}{91.6\%} (24) & \textcolor{red}{87.5\%} (24) & \textcolor{red}{87.5\%} (24) \\
& \textcolor{blue}{100\%} (25) & \textcolor{blue}{100\%} (24) & \textcolor{blue}{100\%} (22) & \textcolor{blue}{100\%} (24) \\
\hline
\multirow{2}{*}{Local$\epsilon$} & \textcolor{red}{76.0\%} (25) & \textcolor{red}{91.6\%} (24) & \textcolor{red}{95.8\%} (24) & \textcolor{red}{91.3\%} (23) \\
& \textcolor{blue}{100\%} (25) & \textcolor{blue}{100\%} (24) & \textcolor{blue}{100\%} (22) & \textcolor{blue}{100\%} (24) \\
\multirow{2}{*}{Local$\zeta$} & \textcolor{red}{92.0\%} (25) & \textcolor{red}{91.6\%} (24) & \textcolor{red}{95.8\%} (24) & \textcolor{red}{91.3\%} (23) \\
& \textcolor{blue}{100\%} (25) & \textcolor{blue}{100\%} (24) & \textcolor{blue}{100\%} (22) & \textcolor{blue}{100\%} (24) \\
\hline
\end{tabular}
\end{center}
\end{table}

\begin{table}[p]
\caption{Classification accuracy of the different gender assignment methods for the authors in South Korea.}
\label{table:s3}
\begin{center}
\begin{tabular}{| D | E | E | E | E |} \hline
\multirow{2}{*}{Method} & \multicolumn{4}{c|}{Year of the first publication} \\ \cline{2-5}
& 1990 or before & 1991--2000 & 2001--2010 & 2011--2020 \\ \hline
\multirow{2}{*}{Global} & \textcolor{red}{75.0\%} (24) & \textcolor{red}{83.3\%} (24) & \textcolor{red}{78.2\%} (23) & \textcolor{red}{100\%} (22) \\
& \textcolor{blue}{100\%} (23) & \textcolor{blue}{100\%} (25) & \textcolor{blue}{95.8\%} (24) & \textcolor{blue}{90.9\%} (22) \\
\multirow{2}{*}{Local} & \textcolor{red}{69.5\%} (23) & \textcolor{red}{75.0\%} (24) & \textcolor{red}{73.9\%} (23) & \textcolor{red}{100\%} (22) \\
& \textcolor{blue}{100\%} (23) & \textcolor{blue}{100\%} (25) & \textcolor{blue}{95.8\%} (24) & \textcolor{blue}{86.3\%} (22) \\
\multirow{2}{*}{Hybrid} & \textcolor{red}{70.8\%} (24) & \textcolor{red}{75.0\%} (24) & \textcolor{red}{73.9\%} (23) & \textcolor{red}{100\%} (23) \\
& \textcolor{blue}{100\%} (23) & \textcolor{blue}{100\%} (25) & \textcolor{blue}{95.8\%} (24) & \textcolor{blue}{90.9\%} (22) \\
\hline
\multirow{2}{*}{Global$\alpha$} & \textcolor{red}{91.3\%} (23) & \textcolor{red}{95.8\%} (24) & \textcolor{red}{95.8\%} (24) & \textcolor{red}{100\%} (25) \\
& \textcolor{blue}{100\%} (22) & \textcolor{blue}{100\%} (24) & \textcolor{blue}{100\%} (24) & \textcolor{blue}{91.6\%} (24) \\
\multirow{2}{*}{Local$\alpha$} & \textcolor{red}{95.6\%} (23) & \textcolor{red}{95.8\%} (24) & \textcolor{red}{91.6\%} (24) & \textcolor{red}{100\%} (25) \\
& \textcolor{blue}{100\%} (23) & \textcolor{blue}{100\%} (25) & \textcolor{blue}{100\%} (24) & \textcolor{blue}{91.6\%} (24) \\
\multirow{2}{*}{Hybrid$\alpha$} & \textcolor{red}{95.6\%} (23) & \textcolor{red}{95.8\%} (24) & \textcolor{red}{91.6\%} (24) & \textcolor{red}{100\%} (25) \\
& \textcolor{blue}{100\%} (23) & \textcolor{blue}{100\%} (25) & \textcolor{blue}{100\%} (24) & \textcolor{blue}{87.5\%} (24) \\
\hline
\end{tabular}
\end{center}
\end{table}

\begin{table}[p]
\caption{Classification accuracy of the different gender assignment methods for the authors in the other countries.}
\label{table:s4}
\begin{center}
\begin{tabular}{| D | E | E | E | E |} \hline
\multirow{2}{*}{Method} & \multicolumn{4}{c|}{Year of the first publication} \\ \cline{2-5}
& 1990 or before & 1991--2000 & 2001--2010 & 2011--2020 \\ \hline
\multirow{2}{*}{Global} & \textcolor{red}{91.6\%} (24) & \textcolor{red}{96.0\%} (25) & \textcolor{red}{100\%} (24) & \textcolor{red}{100\%} (25) \\
& \textcolor{blue}{100\%} (25) & \textcolor{blue}{100\%} (25) & \textcolor{blue}{100\%} (22) & \textcolor{blue}{100\%} (21) \\
\multirow{2}{*}{Local} & \textcolor{red}{79.1\%} (24) & \textcolor{red}{95.8\%} (24) & \textcolor{red}{86.9\%} (23) & \textcolor{red}{96.0\%} (25) \\
& \textcolor{blue}{100\%} (25) & \textcolor{blue}{100\%} (25) & \textcolor{blue}{100\%} (22) & \textcolor{blue}{100\%} (20) \\
\multirow{2}{*}{Hybrid} & \textcolor{red}{79.1\%} (24) & \textcolor{red}{95.8\%} (24) & \textcolor{red}{86.9\%} (23) & \textcolor{red}{96.0\%} (25) \\
& \textcolor{blue}{100\%} (25) & \textcolor{blue}{100\%} (25) & \textcolor{blue}{100\%} (22) & \textcolor{blue}{100\%} (21) \\
\hline
\multirow{2}{*}{Global$\alpha$} & \textcolor{red}{91.6\%} (24) & \textcolor{red}{96.0\%} (25) & \textcolor{red}{100\%} (24) & \textcolor{red}{100\%} (25) \\
& \textcolor{blue}{100\%} (25) & \textcolor{blue}{100\%} (25) & \textcolor{blue}{100\%} (22) & \textcolor{blue}{100\%} (21) \\
\multirow{2}{*}{Local$\alpha$} & \textcolor{red}{79.1\%} (24) & \textcolor{red}{100\%} (24) & \textcolor{red}{86.9\%} (23) & \textcolor{red}{96.0\%} (25) \\
& \textcolor{blue}{100\%} (25) & \textcolor{blue}{100\%} (25) & \textcolor{blue}{100\%} (22) & \textcolor{blue}{100\%} (21) \\
\multirow{2}{*}{Hybrid$\alpha$} & \textcolor{red}{79.1\%} (24) & \textcolor{red}{100\%} (24) & \textcolor{red}{86.9\%} (23) & \textcolor{red}{96.0\%} (25) \\
& \textcolor{blue}{100\%} (25) & \textcolor{blue}{100\%} (25) & \textcolor{blue}{100\%} (22) & \textcolor{blue}{100\%} (21) \\
\hline
\end{tabular}
\end{center}
\end{table}

\begin{table}[p]
\caption{The number of authors by research discipline and country. In each cell, we report the number of gender-assigned authors and the fraction of female authors.}
\label{table:s5}
\begin{center}
\begin{tabular}{| G | H | H | H | H |} \hline
\multirow{2}{*}{Discipline} & \multicolumn{4}{c|}{Country} \\ \cline{2-5}
 & China & Japan & South Korea & Other countries \\ \hline
Art & 568 (25.7\%) & 981 (21.7\%) & 191 (28.6\%) & 206,721 (47.3\%) \\ \hline
Biology & 73,842 (20.7\%) & 221,154 (29.9\%) & 49,992 (27.3\%) & 3,051,331 (46.0\%) \\ \hline
Business & 2,471 (16.7\%) & 2,565 (26.5\%) & 1,643 (15.9\%) & 157,736 (38.2\%) \\ \hline
Chemistry & 46,771 (14.4\%) & 185,064 (26.3\%) & 30,485 (17.8\%) & 1,240,700 (35.8\%) \\ \hline
Computer science & 60,372 (7.4\%) & 92,593 (16.3\%) & 39,646 (8.1\%) & 1,299,940 (23.1\%) \\ \hline
Economics & 4,356 (12.7\%) & 6,215 (21.6\%) & 2,517 (13.8\%) & 266,161 (29.9\%) \\ \hline
Engineering & 18,992 (4.3\%) & 33,866 (11.8\%) & 13,854 (4.9\%) & 389,108 (17.8\%) \\ \hline
Environmental science & 327 (7.6\%) & 1,153 (22.3\%) & 119 (13.3\%) & 19,905 (33.2\%) \\ \hline
Geography & 1,672 (11.6\%) & 4,207 (22.7\%) & 415 (15.3\%) & 94,890 (38.0\%) \\ \hline
Geology & 7,971 (7.3\%) & 16,838 (14.5\%) & 2,377 (11.4\%) & 245,045 (25.7\%) \\ \hline
History & 406 (14.4\%) & 933 (25.4\%) & 153 (20.8\%) & 64,286 (33.9\%) \\ \hline
Materials science & 25,062 (6.8\%) & 76,712 (16.8\%) & 29,290 (6.6\%) & 397,460 (24.3\%) \\ \hline
Mathematics & 12,509 (7.9\%) & 17,798 (21.1\%) & 4,603 (7.1\%) & 293,951 (21.5\%) \\ \hline
Medicine & 66,683 (14.5\%) & 377,948 (29.2\%) & 113,489 (19.7\%) & 5,096,696 (42.9\%) \\ \hline
Philosophy & 1,039 (20.8\%) & 2,476 (26.8\%) & 384 (20.2\%) & 187,240 (38.7\%) \\ \hline
Physics & 31,763 (6.0\%) & 134,033 (15.3\%) & 27,085 (6.2\%) & 843,370 (17.6\%) \\ \hline
Political science & 2,975 (14.8\%) & 4,092 (22.6\%) & 1,213 (21.8\%) & 391,284 (40.2\%) \\ \hline
Psychology & 4,687 (32.2\%) & 17,517 (38.8\%) & 3,647 (34.5\%) & 635,784 (55.7\%) \\ \hline
Sociology & 468 (26.8\%) & 1,155 (37.2\%) & 209 (33.4\%) & 109,660 (57.2\%) \\ \hline
\end{tabular}
\end{center}
\end{table}

\begin{table}[p]
\caption{Statistical significance of the gender gap in the research career for the authors in China. We did not run the $t$-test for the cells with `N/A' because all authors in the group have the same value of the indicator.}
\label{table:s6}
\begin{center}
\begin{tabular}{| F | D | K | K | L | L | G |} \hline
Indicator & Group & $\mu_{\text{F}}$ & $\mu_{\text{M}}$ & $t$-statistic & $p$-value & 95\% CI of $\mu_{\text{F}} - \mu_{\text{M}}$ \\ \hline
\multirow{4}{*}{Productivity} & Overall & 4.47 & 4.93 & $-4.27$ & $< 0.00100$ & ($-0.670$, $-0.248$) \\
& Top 20\% & 10.4 & 12.7 & $-5.36$ & $< 0.00100$ & ($-3.24$, $-1.50$) \\
& Middle 20\% & 3.19 & 3.13 & $3.62$ & $< 0.00100$ & ($0.0263$, $0.0886$) \\
& Low 20\% & 2.00 & 2.00 & N/A & N/A & N/A \\
\hline
\multirow{4}{*}{Total impact} & Overall & 10.1 & 7.31 & $6.80$ & $< 0.00100$ & ($1.96$, $3.55$) \\
& Top 20\% & 34.4 & 28.5 & $3.48$ & $< 0.00100$ & ($2.54$, $9.11$) \\
& Middle 20\% & 4.47 & 1.97 & $67.3$ & $< 0.00100$ & ($2.42$, $2.57$) \\
& Low 20\% & 0.414 & 0.0134 & $32.8$ & $< 0.00100$ & ($0.376$, $0.424$) \\
\hline
\multirow{4}{*}{Annual productivity} & Overall & 1.74 & 1.69 & $2.52$ & $0.0123$ & ($0.0123$, $0.0989$) \\
& Top 20\% & 3.57 & 3.48 & $1.57$ & $0.116$ & ($-0.0203$, $0.185$) \\
& Middle 20\% & 1.51 & 1.46 & $8.04$ & $< 0.00100$ & ($0.0335$, $0.0551$) \\
& Low 20\% & 0.550 & 0.510 & $4.74$ & $< 0.00100$ & ($0.0234$, $0.0565$) \\
\hline
\multirow{4}{*}{Career length} & Overall & 3.24 & 3.68 & $-4.99$ & $< 0.00100$ & ($-0.603$, $-0.263$) \\
& Top 20\% & 9.79 & 11.3 & $-5.22$ & $< 0.00100$ & ($-2.08$, $-0.945$) \\
& Middle 20\% & 1.78 & 1.93 & $-9.33$ & $< 0.00100$ & ($-0.182$, $-0.119$) \\
& Low 20\% & 1.00 & 1.00 & N/A & N/A & N/A \\
\hline
\end{tabular}
\end{center}
\end{table}

\begin{table}[p]
\caption{Statistical significance of the gender gap in the research career for the authors in Japan. }
\label{table:s7}
\begin{center}
\begin{tabular}{| F | D | K | K | L | L | G |} \hline
Indicator & Group & $\mu_{\text{F}}$ & $\mu_{\text{M}}$ & $t$-statistic & $p$-value & 95\% CI of $\mu_{\text{F}} - \mu_{\text{M}}$ \\ \hline
\multirow{4}{*}{Productivity} & Overall & 5.76 & 10.0 & $-81.9$ & $< 0.00100$ & ($-4.37$, $-4.16$) \\
& Top 20\% & 15.0 & 32.5 & $-84.0$ & $< 0.00100$ & ($-18.0$, $-17.1$) \\
& Middle 20\% & 3.61 & 4.61 & $-172$ & $< 0.00100$ & ($-1.01$, $-0.984$) \\
& Low 20\% & 2.00 & 2.00 & N/A & N/A & N/A \\
\hline
\multirow{4}{*}{Total impact} & Overall & 8.16 & 10.2 & $-19.0$ & $< 0.00100$ & ($-2.24$, $-1.82$) \\
& Top 20\% & 29.2 & 40.2 & $-24.6$ & $< 0.00100$ & ($-11.9$, $-10.1$) \\
& Middle 20\% & 3.16 & 2.75 & $52.5$ & $< 0.00100$ & ($0.398$, $0.428$) \\
& Low 20\% & 0.239 & 0.128 & $45.7$ & $< 0.00100$ & ($0.107$, $0.116$) \\
\hline
\multirow{4}{*}{Annual productivity} & Overall & 1.36 & 1.32 & $7.60$ & $< 0.00100$ & ($0.0281$, $0.0477$) \\
& Top 20\% & 2.85 & 2.91 & $-4.62$ & $< 0.00100$ & ($-0.0804$, $-0.0325$) \\
& Middle 20\% & 1.14 & 1.07 & $52.2$ & $< 0.00100$ & ($0.0654$, $0.0705$) \\
& Low 20\% & 0.403 & 0.340 & $42.9$ & $< 0.00100$ & ($0.0606$, $0.0664$) \\
\hline
\multirow{4}{*}{Career length} & Overall & 5.15 & 8.53 & $-107$ & $< 0.00100$ & ($-3.44$, $-3.32$) \\
& Top 20\% & 14.0 & 24.3 & $-126$ & $< 0.00100$ & ($-10.4$, $-10.0$) \\
& Middle 20\% & 3.28 & 4.88 & $-235$ & $< 0.00100$ & ($-1.61$, $-1.59$) \\
& Low 20\% & 1.00 & 1.14 & $-96.6$ & $< 0.00100$ & ($-0.145$, $-0.139$) \\
\hline
\end{tabular}
\end{center}
\end{table}

\begin{table}[p]
\caption{Statistical significance of the gender gap in the research career for the authors in South Korea. }
\label{table:s8}
\begin{center}
\begin{tabular}{| F | D | K | K | L | L | G |} \hline
Indicator & Group & $\mu_{\text{F}}$ & $\mu_{\text{M}}$ & $t$-statistic & $p$-value & 95\% CI of $\mu_{\text{F}} - \mu_{\text{M}}$ \\ \hline
\multirow{4}{*}{Productivity} & Overall & 4.92 & 6.17 & $-17.0$ & $< 0.00100$ & ($-1.39$, $-1.10$) \\
& Top 20\% & 11.5 & 16.7 & $-20.2$ & $< 0.00100$ & ($-5.74$, $-4.73$) \\
& Middle 20\% & 3.50 & 3.67 & $-11.1$ & $< 0.00100$ & ($-0.204$, $-0.142$) \\
& Low 20\% & 2.00 & 2.00 & N/A & N/A & N/A \\
\hline
\multirow{4}{*}{Total impact} & Overall & 7.45 & 6.13 & $5.95$ & $< 0.00100$ & ($0.888$, $1.76$) \\
& Top 20\% & 25.4 & 22.4 & $3.29$ & $0.00102$ & ($1.20$, $4.75$) \\
& Middle 20\% & 3.25 & 2.20 & $50.2$ & $< 0.00100$ & ($1.01$, $1.09$) \\
& Low 20\% & 0.276 & 0.137 & $19.3$ & $< 0.00100$ & ($0.125$, $0.153$) \\
\hline
\multirow{4}{*}{Annual productivity} & Overall & 1.54 & 1.42 & $8.01$ & $< 0.00100$ & ($0.0955$, $0.157$) \\
& Top 20\% & 3.24 & 3.06 & $4.62$ & $< 0.00100$ & ($0.106$, $0.263$) \\
& Middle 20\% & 1.31 & 1.17 & $33.8$ & $< 0.00100$ & ($0.134$, $0.151$) \\
& Low 20\% & 0.448 & 0.385 & $13.2$ & $< 0.00100$ & ($0.0539$, $0.0726$) \\
\hline
\multirow{4}{*}{Career length} & Overall & 3.91 & 5.27 & $-22.9$ & $< 0.00100$ & ($-1.48$, $-1.25$) \\
& Top 20\% & 10.3 & 14.2 & $-24.5$ & $< 0.00100$ & ($-4.24$, $-3.61$) \\
& Middle 20\% & 2.51 & 3.34 & $-51.9$ & $< 0.00100$ & ($-0.859$, $-0.796$) \\
& Low 20\% & 1.00 & 1.00 & N/A & N/A & N/A \\
\hline
\end{tabular}
\end{center}
\end{table}

\begin{table}[p]
\caption{Statistical significance of the gender gap in the research career for the authors in the other countries. }
\label{table:s9}
\begin{center}
\begin{tabular}{| F | D | K | K | L | L | G |} \hline
Indicator & Group & $\mu_{\text{F}}$ & $\mu_{\text{M}}$ & $t$-statistic & $p$-value & 95\% CI of $\mu_{\text{F}} - \mu_{\text{M}}$ \\ \hline
\multirow{4}{*}{Productivity} & Overall & 5.73 & 8.33 & $-172$ & $< 0.00100$ & ($-2.63$, $-2.57$) \\
& Top 20\% & 15.1 & 26.1 & $-178$ & $< 0.00100$ & ($-11.1$, $-10.9$) \\
& Middle 20\% & 3.54 & 4.08 & $-336$ & $< 0.00100$ & ($-0.541$, $-0.535$) \\
& Low 20\% & 2.00 & 2.00 & N/A & N/A & N/A \\
\hline
\multirow{4}{*}{Total impact} & Overall & 14.3 & 18.1 & $-57.5$ & $< 0.00100$ & ($-3.88$, $-3.62$) \\
& Top 20\% & 52.5 & 72.0 & $-67.2$ & $< 0.00100$ & ($-20.1$, $-18.9$) \\
& Middle 20\% & 5.19 & 4.76 & $127$ & $< 0.00100$ & ($0.426$, $0.440$) \\
& Low 20\% & 0.431 & 0.310 & $118$ & $< 0.00100$ & ($0.120$, $0.124$) \\
\hline
\multirow{4}{*}{Annual productivity} & Overall & 1.20 & 1.18 & $16.2$ & $< 0.00100$ & ($0.0174$, $0.0222$) \\
& Top 20\% & 2.56 & 2.65 & $-29.0$ & $< 0.00100$ & ($-0.0936$, $-0.0818$) \\
& Middle 20\% & 0.996 & 0.947 & $155$ & $< 0.00100$ & ($0.0484$, $0.0496$) \\
& Low 20\% & 0.335 & 0.281 & $162$ & $< 0.00100$ & ($0.0531$, $0.0544$) \\
\hline
\multirow{4}{*}{Career length} & Overall & 6.16 & 8.75 & $-240$ & $< 0.00100$ & ($-2.61$, $-2.57$) \\
& Top 20\% & 17.3 & 25.6 & $-314$ & $< 0.00100$ & ($-8.28$, $-8.17$) \\
& Middle 20\% & 3.70 & 4.86 & $-539$ & $< 0.00100$ & ($-1.16$, $-1.16$) \\
& Low 20\% & 1.00 & 1.18 & $-260$ & $< 0.00100$ & ($-0.174$, $-0.171$) \\
\hline
\end{tabular}
\end{center}
\end{table}

\begin{table}[p]
\caption{Pearson correlation coefficients between the career length and the number of coauthors of all the authors, female authors, and male authors in each country group.}
\label{table:s10}
\begin{center}
\begin{tabular}{| I | J | J | J |} \hline 
\multirow{2}{*}{Country} & \multicolumn{3}{c|}{Gender} \\ \cline{2-4}
 & All & Female & Male \\ \hline
China & 0.35 & 0.32 & 0.36 \\ \hline
Japan & 0.49 & 0.39 & 0.49 \\ \hline
South Korea & 0.45 & 0.30 & 0.45 \\ \hline
Other countries & 0.40 & 0.33 & 0.41 \\ \hline
\end{tabular}
\end{center}
\end{table}

\newpage

\renewcommand{\refname}{Supplementary References}

\end{document}